\documentclass[zpreprint,zbstdefault]{./LaTeX/zeus/zeus_paper}
\usepackage[english]{babel}
\usepackage{./LaTeX/styles/lineno}
\usepackage{longtable,lscape}

\chardef\usc=95
\chardef\til=126
\catcode`\@=11 
\DeclareRobustCommand\xdotspace{\futurelet\@let@token\@xdotspace}
\def\@xdotspace{%
  \ifx\@let@token.\else
  \ifx\@let@token\bgroup.\else
  \ifx\@let@token\egroup.\else
  \ifx\@let@token\/.\else
  \ifx\@let@token\ .\else
  \ifx\@let@token~.\else
  \ifx\@let@token!.\else
  \ifx\@let@token,.\else
  \ifx\@let@token:.\else
  \ifx\@let@token;.\else
  \ifx\@let@token?.\else
  \ifx\@let@token/.\else
  \ifx\@let@token'.\else
  \ifx\@let@token).\else
  \ifx\@let@token-.\else
  \ifx\@let@token\@xobeysp.\else
  \ifx\@let@token\space.\else
  \ifx\@let@token\@sptoken.\else
   .\space
   \fi\fi\fi\fi\fi\fi\fi\fi\fi\fi\fi\fi\fi\fi\fi\fi\fi\fi}
\catcode`\@=12 

\newcommand{\stru}[2]{%
   \relax\ifmmode\hbox{\vrule height#1 depth#2 width0pt}%
   \else\vrule height#1 depth#2 width0pt\fi}

\newcommand{\Ronum}[1]{\uppercase\expandafter{\romannumeral#1}}
\newcommand{\ronum}[1]{\expandafter{\romannumeral#1}}
\DeclareRobustCommand{\LaTeXZ}{%
  \LaTeX\kern-.05em4\kern-.1em
  {\raisebox{-0.2ex}{$\scriptstyle\text{ZEUS}$}}\xspace}

\DeclareMathAlphabet{\mathbf}{OT1}{cmr}{bx}{sl}
\newcommand{\eVdist}{\kern-0.06667em}

\newcommand{\gev}{{\,\text{Ge}\eVdist\text{V\/}}}

\newcommand{\Tesla}{\,\text{T}}

\newcommand{\slashfrac}[2]{%
  \raisebox{0.5ex}{\ensuremath #1}\kern-0.12em/\kern-0.08em
  \raisebox{-.8ex}{\ensuremath #2}}

\newcommand{\sqr}[3]{%
    {\vcenter{\hrule height.#3ex\hbox{\vrule width.#2ex height#1ex
     \kern#1ex\vrule width.#3ex}\hrule height.#2ex}}}

\catcode`\@=11 
\newcommand{\parenbar}{\mathpalette\p@renb@r}
\def\p@renb@r#1#2{\vbox{%
  \ifx#1\scriptscriptstyle \dimen@.7em\dimen@ii.2em\else
  \ifx#1\scriptstyle \dimen@.8em\dimen@ii.25em\else
  \dimen@1em\dimen@ii.4em\fi\fi \offinterlineskip
  \ialign{\hfill##\hfill\cr
    \vbox{\hrule width\dimen@ii}\cr
    \noalign{\vskip-.3ex}%
    \hbox to\dimen@{$\mathchar300\hfil\mathchar301$}\cr
    \noalign{\vskip-.3ex}%
    $#1#2$\cr}}}
\catcode`\@=12 

\newcommand{\IP}{{\rm I$\kern-0.01667em$P}\xspace}

\mathchardef\qsm=63
\mathchardef\pls=43
\mathchardef\mns=512
\mathchardef\plm=518
\mathchardef\eql=61
\mathchardef\smallleft=300
\mathchardef\smallright=301
\mathchardef\les=316
\mathchardef\gre=318
\mathchardef\leq=532
\mathchardef\grq=533

\catcode`\@=11 
\newcounter{pict@width}
\newcounter{pict@height}
\newlength{\pict@scale}
\newcommand{\psfigadd}[4]{%
\setcounter{pict@width}{1*\ratio{#2+\pict@scale/2}{\pict@scale}}
\setcounter{pict@height}{1*\ratio{#3+\pict@scale/2}{\pict@scale}}
\setlength{\unitlength}{\pict@scale}
\hbox to #2{\hspace{-\fill}\begin{picture}(\thepict@width,\thepict@height)
\put(0,0){\psfig{figure=#1,width=#2,height=#3,clip=}}
\SetScale{0.283466457}
\SetWidth{1.763889}
{#4}
\end{picture}}
}
\newcounter{pict@widthfst}
\newcounter{pict@widthscd}
\newcounter{pict@widthtot}
\newcommand{\psfigaddtwo}[7]{%
\setcounter{pict@widthfst}{1*\ratio{#2+\pict@scale/2}{\pict@scale}}
\setcounter{pict@widthscd}{1*\ratio{#2+#4+\pict@scale/2}{\pict@scale}}
\setcounter{pict@widthtot}{1*\ratio{#2+#4+#6+\pict@scale/2}{\pict@scale}}
\setcounter{pict@height}{1*\ratio{#3+\pict@scale/2}{\pict@scale}}
\setlength{\unitlength}{\pict@scale}
\hbox{\hspace{-\fill}\begin{picture}(\thepict@widthtot,\thepict@height)
\put(0,0){\psfig{figure=#1,width=#2,height=#3,clip=}}
\put(\thepict@widthscd,0){\psfig{figure=#5,width=#6,height=#3,clip=}}
\SetScale{0.283466457}
\SetWidth{1.763889}
{#7}
\end{picture}}
}
\newcommand{\psfigror}[4]{%
\setcounter{pict@width}{1*\ratio{#2+\pict@scale/2}{\pict@scale}}
\setcounter{pict@height}{1*\ratio{#3+\pict@scale/2}{\pict@scale}}
\setlength{\unitlength}{\pict@scale}
\hbox{\begin{picture}(\thepict@width,\thepict@height)
\put(0,\thepict@height){\psfig{figure=#1,width=#3,height=#2,clip=,angle=270}}
\SetScale{0.283466457}
\SetWidth{1.763889}
{#4}
\end{picture}}
}
\newcommand{\psfigrol}[4]{%
\setcounter{pict@width}{1*\ratio{#2+\pict@scale/2}{\pict@scale}}
\setcounter{pict@height}{1*\ratio{#3+\pict@scale/2}{\pict@scale}}
\setlength{\unitlength}{\pict@scale}
\hbox{\begin{picture}(\thepict@width,\thepict@height)
\put(0,0){\psfig{figure=#1,width=#3,height=#2,clip=,angle=90}}
\SetScale{0.283466457}
\SetWidth{1.763889}
{#4}
\end{picture}}
}
\catcode`\@=12 
\newlength\listtextwidth

\catcode`\@=11 
\newlength{\@tabfninsert}
\newlength{\@tabfnwidth}
\newcommand{\tabfootnote}[2]{%
  \setlength{\@tabfninsert}{0.8em}
  \setlength{\@tabfnwidth}{\textwidth}
  \addtolength{\@tabfnwidth}{-\@tabfninsert}
  \addtolength{\@tabfnwidth}{-0.4em}
  \noindent\makebox[\@tabfninsert][r]{\footnotesize$^{#1}$\hfil}\hfill%
  \parbox[t]{\@tabfnwidth}{\footnotesize #2\hfill}}
\catcode`\@=12 

\def\citeCTD{{\cite{%
nim:a279:290,*npps:b32:181,*nim:a338:254%
}}\xspace}
\def\citeMVD{{\cite{%
nim:a581:656%
}}\xspace}
\def\citeCAL{{\cite{%
nim:a309:77,*nim:a309:101,*nim:a321:356,*nim:a336:23%
}}\xspace}

\def\citePCAL{{\cite{%
desy-92-066,*zfp:c63:391,*acpp:b32:2025%
}}\xspace}

\begin{document}
\prepnum{DESY--09--046}

\title{
Measurement of the longitudinal proton structure function at HERA
}                                                       
                    
\author{ZEUS Collaboration\\}

\date{March 2009}

\abstract{
The reduced cross sections for $ep$ deep inelastic scattering have been 
measured with the ZEUS detector at HERA at three different centre-of-mass 
energies, 318, 251 and 225\gev. From the 
cross sections, measured double differentially in Bjorken $x$ and 
the virtuality, $Q^2$, the proton structure functions $F_L$ and $F_2$ 
have been extracted in the region $5\times 10^{-4}<x<0.007$ and $20<Q^2<130$\gev$^2$. 
}
\makezeustitle

\pagenumbering{Roman}                                                                              
\begin{center}                                                                                     
{                      \Large  The ZEUS Collaboration              }                               
\end{center}                                                                                       
  S.~Chekanov,                                                                                     
  M.~Derrick,                                                                                      
  S.~Magill,                                                                                       
  B.~Musgrave,                                                                                     
  D.~Nicholass$^{   1}$,                                                                           
  \mbox{J.~Repond},                                                                                
  R.~Yoshida\\                                                                                     
 {\it Argonne National Laboratory, Argonne, Illinois 60439-4815, USA}~$^{n}$                       
\par \filbreak                                                                                     
  M.C.K.~Mattingly \\                                                                              
 {\it Andrews University, Berrien Springs, Michigan 49104-0380, USA}                               
\par \filbreak                                                                                     
  P.~Antonioli,                                                                                    
  G.~Bari,                                                                                         
  L.~Bellagamba,                                                                                   
  D.~Boscherini,                                                                                   
  A.~Bruni,                                                                                        
  G.~Bruni,                                                                                        
  F.~Cindolo,                                                                                      
  M.~Corradi,                                                                                      
\mbox{G.~Iacobucci},                                                                               
  A.~Margotti,                                                                                     
  R.~Nania,                                                                                        
  A.~Polini\\                                                                                      
  {\it INFN Bologna, Bologna, Italy}~$^{e}$
% ...
\newpage                                                                                           
\enlargethispage{5cm}                                                                              
$^{\    1}$ also affiliated with University College London,                                        
United Kingdom\\                                                                                   
$^{\    2}$ now at University of Salerno, Italy \\                                                 
% ...
$^{\    5}$ supported by the research grant No. 1 P03B 04529 (2005-2008) \\                        
% ...
%
\newpage   
                                                           %
                                                           %
\begin{tabular}[h]{rp{14cm}}                                                                       
$^{a}$ &  supported by the Natural Sciences and Engineering Research Council of Canada (NSERC) \\  
$^{b}$ &  supported by the German Federal Ministry for Education and Research (BMBF), under        
          contract Nos. 05 HZ6PDA, 05 HZ6GUA, 05 HZ6VFA and 05 HZ4KHA\\                            
$^{c}$ &  supported in part by the MINERVA Gesellschaft f\"ur Forschung GmbH, the Israel Science   
          Foundation (grant No. 293/02-11.2) and the US-Israel Binational Science Foundation \\    
$^{d}$ &  supported by the Israel Science Foundation\\                                             
$^{e}$ &  supported by the Italian National Institute for Nuclear Physics (INFN) \\                
$^{f}$ &  supported by the Japanese Ministry of Education, Culture, Sports, Science and Technology 
          (MEXT) and its grants for Scientific Research\\                                          
$^{g}$ &  supported by the Korean Ministry of Education and Korea Science and Engineering          
          Foundation\\                                                                             
$^{h}$ &  supported by the Netherlands Foundation for Research on Matter (FOM)\\                   
$^{i}$ &  supported by the Polish State Committee for Scientific Research, project No.             
          DESY/256/2006 - 154/DES/2006/03\\                                                        
$^{j}$ &  partially supported by the German Federal Ministry for Education and Research (BMBF)\\   
$^{k}$ &  supported by RF Presidential grant N 1456.2008.2 for the leading                         
          scientific schools and by the Russian Ministry of Education and Science through its      
          grant for Scientific Research on High Energy Physics\\                                   
$^{l}$ &  supported by the Spanish Ministry of Education and Science through funds provided by     
          CICYT\\                                                                                  
$^{m}$ &  supported by the Science and Technology Facilities Council, UK\\                         
$^{n}$ &  supported by the US Department of Energy\\                                               
$^{o}$ &  supported by the US National Science Foundation. Any opinion,                            
findings and conclusions or recommendations expressed in this material                             
are those of the authors and do not necessarily reflect the views of the                           
National Science Foundation.\\                                                                     
$^{p}$ &  supported by the Polish Ministry of Science and Higher Education                         
as a scientific project (2006-2008)\\                                                              
$^{q}$ &  supported by FNRS and its associated funds (IISN and FRIA) and by an Inter-University    
          Attraction Poles Programme subsidised by the Belgian Federal Science Policy Office\\     
$^{r}$ &  supported by an FRGS grant from the Malaysian government\\                               
\end{tabular}                                                                                      
                                                           %
                                                           %

\pagenumbering{arabic}
\section{Introduction}

The inclusive $e^\pm p$ deep inelastic scattering (DIS) cross section
can, at low virtuality of the exchanged boson, $Q^2$, be expressed in
terms of the two structure functions, $F_2$ and $F_L$, as
\begin{equation}
\frac{d^2\sigma^{e^\pm p}}{dxdQ^2} = \frac{2\pi\alpha^2Y_+}{xQ^4}
\left[F_2(x,Q^2) - \frac{y^2}{Y_+} F_L(x,Q^2)\right]
= \frac{2\pi\alpha^2Y_+}{xQ^4} \; \tilde{\sigma} (x,Q^2,y), 
\label{eq:disnc-xsec}
\end{equation}
where $\alpha$ is the fine structure constant, $x$ is the Bjorken
scaling variable, $y$ is the inelasticity and $Y_+=1+(1-y)^2$. The
quantity $\tilde{\sigma}$ is referred to as the reduced cross
section. The kinematic variables are related via $Q^2=xys$, where $\sqrt{s}$
is the $ep$ centre-of-mass energy.
The magnitude of $F_L$ is proportional to the absorption cross section
of  longitudinally polarised virtual photons by protons,
$F_L\propto \sigma_L$, while $F_2$ includes also the absorption cross section for
transversely polarised virtual photons,
$F_2\propto(\sigma_T+\sigma_L)$. At low values of $x$
and small $\sigma_L$, 
the ratio
$R=F_L/(F_2-F_L)\approx\sigma_L/\sigma_T$ gives the relative
strength of the two components.

HERA measurements of the reduced $ep$ DIS cross section and
$F_2$
~\cite{Chekanov:2001qu,Adloff:2000qk,Collaboration:2009bp,*Collaboration:2009kv} 
provide the strongest
constraints on the proton parton distribution functions (PDFs) at low
$x$. Within the DGLAP formalism~\cite{Art:dokshitzer,*Art:gribov,*Art:gribov2,*Art:altarelli}, $F_2$ at low $x$ is dominated by the
$q\bar{q}$ sea distributions while the scaling violations of $F_2$
reflect the gluon distribution, $g(x,Q^2)$, via a convolution with the
splitting function $P_{qg}(x)$,
${\partial F_2}/{\partial \ln Q^2}\sim \alpha_s(Q^2)P_{qg}(x) \otimes xg(x,Q^2)$, 
where $\alpha_s$ is the strong coupling constant.

The published values of $F_2$ at low $x$ at HERA required assumptions 
to be made about $F_L$ or were restricted to the kinematic region where
the contribution from $F_L$ was sufficiently suppressed to be neglected.  Moreover, gluon distributions extracted from
scaling violations are dependent on 
the formalism~\cite{Catani:1996sc} and the order of perturbative
expansion~\cite{Martin:2006qv} used to calculate the splitting
functions.
Measurements of the reduced cross section at fixed $(x,Q^2)$ and
different $y$ allow $F_2$ and $F_L$ to be extracted simultaneously,
thereby eliminating the assumptions about $F_L$ when extracting
$F_2$. Furthermore, 
a direct
measurement of $F_L$, which is
strongly correlated to the gluon 
density~\cite{Altarelli:1978tq},
provides an important
consistency check of the formalism.

A model-independent determination of $F_L$ requires the reduced 
cross section  to be measured at fixed values of $x$ and $Q^2$ 
for multiple centre-of-mass energies (varying $y$ values). 
This method has been previously used to extract $F_L$ in 
fixed-target experiments~\cite{Aubert:1982ts,Benvenuti:1989rh,Whitlow:1990gk,Arneodo:1996qe} and recently 
by the H1 collaboration~\cite{Aaron:2008tx}. 
The H1 collaboration has also applied extrapolation methods to 
determine $F_L$~\cite{Adloff:2000qk,Adloff:1996yz,*Adloff:2003uh}. 

In this paper, the first 
ZEUS measurement of $F_L$ is presented as well as the most precise 
ZEUS measurement of $F_2$ in the kinematic region studied. 
Comparisons of theoretical predictions with the data are also presented.

%

\section{Experimental method}

The values of $F_2$ and $F_L$ were extracted at fixed $x$ and $Q^2$ by 
fitting a straight line to the values of $\tilde{\sigma}$ against $y^2/Y_{+}$ 
in the so-called Rosenbluth plot~\cite{Rosenbluth:1950yq}.  
The method is based on Eq.~\ref{eq:disnc-xsec}, which implies 
that $F_2(x,Q^2) = \tilde{\sigma}(x,Q^2,y=0)$ and 
$F_L(x,Q^2) = -{\rm \partial}\tilde{\sigma}(x,Q^2,y)/{\rm \partial}(y^2/Y_{+})$, 
which in turn implies 
the need for data at fixed $(x,Q^2)$ and different $y$. At HERA,
this can be achieved by varying $\sqrt{s}$.

The precision of this procedure depends on the 
range available
in $y^2/Y_{+}$. This was maximised by collecting data at the nominal HERA energy, $\sqrt{s}=318$\gev, and at $\sqrt{s}=225$\gev, the lowest attainable energy with adequate instantaneous luminosity. An intermediate data set was collected at $\sqrt{s}=251$\gev.

The variation of $\sqrt{s}$ was achieved by varying the proton beam energy, $E_p$, while keeping the electron beam energy constant, $E_e=27.5$\gev. Data were collected in 2006 and 2007 with $E_p=920$, $575$ and $460$\gev, 
referred to respectively as the 
high- (HER), medium- (MER) and low-energy-running (LER) samples.  
The corresponding integrated luminosities of the HER, MER and LER samples are $44.5$, $7.1$ and $13.9~{\rm pb}^{-1}$, respectively. 



\section{Experimental apparatus}

A detailed description of the ZEUS detector can be found elsewhere~\cite{zeus:1993:bluebook}. A brief outline of the components most relevant for this analysis is given below.

In the kinematic range of the analysis, charged particles were tracked
in the central tracking detector (CTD)~\citeCTD and the microvertex
detector (MVD)~\citeMVD. These components operated in a magnetic field
of $1.43\Tesla$ provided by a thin superconducting solenoid. The CTD
drift chamber, consisting of 72 sense wire layers organised into 9 super layers, covered the polar-angle\footnote{The ZEUS coordinate system is a right-handed Cartesian system, with the $Z$ axis pointing in the proton beam direction, referred to as the ``forward direction'', and the $X$ axis pointing towards the centre of HERA. The coordinate origin is at the nominal interaction point.\xspace} region \mbox{$15^\circ<\theta<164^\circ$}. The MVD silicon tracker consisted of a barrel (BMVD) and a forward (FMVD) section. The BMVD provided polar-angle coverage for tracks with three measurements from $30^\circ$ to $150^\circ$. The FMVD extended the polar-angle coverage in the forward region to $7^\circ$. 

The high-resolution uranium--scintillator calorimeter (CAL)~\citeCAL consisted of three parts: the forward (FCAL), the barrel (BCAL) and the rear (RCAL) calorimeters. Each part was subdivided transversely into towers and longitudinally into one electromagnetic section (EMC) and either one (in RCAL) or two (in BCAL and FCAL) hadronic sections (HAC). The smallest subdivision of the calorimeter was called a cell.  The CAL energy resolutions, as measured under test-beam conditions, were $\sigma(E)/E=0.18/\sqrt{E}$ for electrons and $\sigma(E)/E=0.35/\sqrt{E}$ for hadrons, with $E$ in GeV.

%
The rear hadron-electron separator (RHES)~\cite{Dwurazny:1988fj} consisted of a layer of approximately $10\,000$ 
$(3\times 3\,{\rm cm^2})$ 
silicon-pad detectors inserted in the RCAL at a depth 
of approximately 3 radiation 
lengths. The polar-angle coverage is approximately $131^\circ<\theta<173^\circ$.
The small-angle rear tracking detector
(SRTD)~\cite{Bamberger:1997fg} was attached to the front face of the
RCAL and consisted of two square planes of scintillator strips.  The
detector covers the total area of 68~cm$\times$68~cm, with a
20~cm$\times$20~cm cutout in the centre for the beam-pipe.  The
polar-angle coverage is $162^\circ<\theta<176^\circ$, with full
acceptance for $167^\circ<\theta<174.5^\circ$.

The small tungsten--scintillator calorimeter located approximately 6~m from the interaction point in the rear direction was referred to as the ``6m-tagger''~\cite{thesis:gosau,*thesis:schroder}. For scattered electrons in the energy range from $4.1$ to $7.4$\gev, the acceptance was very close to one with very high purity.

The luminosity was measured using the Bethe-Heitler reaction $ep\,\rightarrow\, e\gamma p$ by a luminosity detector which consisted of an independent lead--scintillator calorimeter\citePCAL and a magnetic spectrometer\cite{Helbich:2005qf} system. The fractional systematic uncertainty on the measured luminosity was 2.6\%.

\section{Event reconstruction and selection}

The kinematic region studied spanned 
$0.09<y<0.78$ 
and $20<Q^2<130$\gev$^2$, corresponding to 
$5\times 10^{-4}<x<0.007$.
The event kinematics were evaluated based on the reconstruction of the scattered electron~\cite{emethod} using
\begin{equation}
y_e = 1 - \frac{E^{\prime}_e}{2E_e}\left(1-\cos\theta_e\right),
\label{eq:ye}
\end{equation}
\begin{equation}
Q_e^2 = 2E^{\prime}_eE_e\left(1+\cos\theta_e\right),
\label{eq:Q2}
\end{equation}
where $\theta_e$ and $E^{\prime}_e$ are the polar angle and energy of the scattered electron, respectively.

Electrons were identified using a neural network based on the moments
of the three-dimensional shower profile of clusters found in the
CAL\cite{Sinkus:1996ch,*Abramowicz:1995zi}. The quantity
$E^{\prime}_e$ was reconstructed using the CAL, and $\theta_e$ was
determined using the reconstructed interaction vertex and scattered
electron position in the SRTD or, if outside the SRTD acceptance, in
the RHES. In less than $2\%$ of events, $\theta_e$ could not
reliably be determined using the SRTD+HES system and such events were rejected.

The quantity
%
\mbox{$\delta \equiv \sum_{i}(E-p_Z)_i$} was used both in the trigger and in the offline 
analysis.
The sum runs over all CAL energy deposits. Conservation of energy, $E$, and 
longitudinal momentum, $p_Z$, implies that $\delta = 2E_e = 55$\gev. Undetected 
particles that escape through the forward beam-pipe hole 
contribute negligibly to~$\delta$. Undetected particles that escape through the 
rear beam-pipe hole, such as the final-state electron in a photoproduction event, 
cause a substantial reduction in $\delta$. Events not originating from $ep$ 
collisions often exhibit a very large $\delta$. 

A three-level trigger system was used to select events online~\cite{zeus:1993:bluebook,epj:c1:109,nim:a355:278,Gttsmith1}. A dedicated trigger was developed providing high efficiency for high-$y$ events~\cite{thesis:shima}. 
The trigger required an event to have $\delta>30$\gev~and either an electron
candidate with $E_e^\prime>4$\gev~in the RCAL and outside a
30~cm$\times$30~cm square centred around the beam-pipe, or
$\delta^{\theta<165^\circ}>20$\gev, where $\delta^{\theta<165^\circ}$
denotes $\delta$ calculated only from the CAL energy deposits at polar
angles less than $165^\circ$.

Events were selected offline if:

\begin{itemize}
\item $42 < \delta < 65$\gev;
\item the reconstructed interaction vertex fulfilled $|Z_{\rm vtx}|<30$~cm;
\item the energy of the most probable electron candidate satisfied $E^{\prime}_e>6$\gev;
\item the event topology was not compatible with an elastic QED Compton (QEDC) event;
\item the event timing was consistent with the HERA bunch structure;
\item $y_e < 0.95$ and $y_{\rm JB} > 0.05$, where $y_{\rm JB}$ is the Jacquet--Blondel estimator~\cite{proc:epfacility:1979:391} of $y$;
\item $p_{T,h}/p_{T,e}>0.3$, where $p_{T,h}$ and $p_{T,e}$ refer to the transverse momentum of the hadronic system and electron candidate, respectively. 
\end{itemize}

The projected path of the electron candidate was required to:
%
\begin{itemize}
\item exit the CTD at a radius $>20$~cm and hence traverse the MVD fiducial volume and at least four CTD sense-wire layers, ensuring the possibility of identifying the track;
\item enter the RCAL at a radius $<135$~cm, missing the inactive region between the RCAL and BCAL sections.
\end{itemize}

Hit information from the MVD and CTD was used to identify the tracks 
of the electron candidates. The procedure was based on the ratios of 
the number of observed to the maximum number of possible hits in the 
MVD and CTD, denoted $f_{\rm hit}^{\rm MVD}$ and $f_{\rm hit}^{\rm CTD}$, 
respectively:
\begin{itemize}
\item $f_{\rm hit}^{\rm MVD}>0.45$; 
\item $f_{\rm hit}^{\rm CTD}>0.6$.
\end{itemize}

This method was used because of the wider polar angular acceptance
compared to the regular tracking capability of the MVD+CTD tracking
system. Specifically, for an event with a nominally placed vertex,
the electron candidate can be validated up to an angle of
$\theta_e=169~\circ$, compared to $\theta_e=159~\circ$ with full tracking.
After all cuts the HER, MER and LER samples contained 819168, 
115719 and 205967 events, respectively.

\section{Cross section determination}
\label{sec:unfold}

The reduced cross sections in a given $(x,Q^2)$ bin were calculated according to
\begin{displaymath}
  \tilde{\sigma}(x,Q^2) =
  \frac{N_\mathrm{data}-N_\mathrm{MC}^\mathrm{bg}}{N_\mathrm{MC}^\mathrm{DIS}}
  \, \tilde{\sigma}_\mathrm{SM}(x,Q^2),
\end{displaymath}
where $\tilde{\sigma}_\mathrm{SM}(x,Q^2)$ is the Standard Model electroweak 
Born-level reduced cross section and $N_\mathrm{data}$, 
$N_\mathrm{MC}^\mathrm{bg}$ and $N_\mathrm{MC}^\mathrm{DIS}$ denote, 
respectively, the number of observed events in the data and the expected 
number of background and DIS events from the Monte Carlo (MC).  
The CTEQ5D~\cite{Art:cteq5l} parameterisation of the proton PDF was used 
when calculating $\tilde{\sigma}_\mathrm{SM}(x,Q^2)$ as well as in the 
MC models when evaluating $N_\mathrm{MC}^\mathrm{DIS}$ and 
$N_\mathrm{MC}^\mathrm{bg}$. Specifically, the DIS signal processes were 
simulated using the {\sc Djangoh 1.6}~\cite{django} MC model.  
After the full event selection, the background consisted almost entirely 
of photoproduction events. These were simulated using 
the {\sc Pythia 6.221}~\cite{Art:pythia, *Art:pythia2} MC model.
The additional background components that were considered 
were elastic QEDC and mis-reconstructed low-$Q^2$ DIS, simulated using 
the {\sc Grape-Compton}~\cite{Abe:2000cv} and {\sc Djangoh 1.6} MC models, 
respectively. The MC events were processed through a full simulation of the 
ZEUS detector and trigger system based on {\sc Geant 3.21}~\cite{unp:geant}. 

The {\sc Djangoh} and {\sc Pythia} samples included a diffractive component 
and first-order electroweak corrections. The diffractive and non-diffractive 
components of the {\sc Djangoh} sample were scaled to improve the description 
of the HER, MER and LER $\eta_{\rm max}$ distributions, where $\eta_{\rm max}$ is equal to the 
pseudorapidity of the most forward CAL energy deposit.  
The electroweak corrections were simulated using 
the {\sc Heracles 4.6}~\cite{django,heracles4.6} MC model. 
Their uncertainty was evaluated by 
comparing the predictions from {\sc Heracles} to the higher-order predictions 
from {\sc Hector 1.0}~\cite{Arbuzov:1995id}.  The predictions were found to 
agree to within $0.5\%$. The hadronic final state of the {\sc Djangoh} MC was simulated using the colour-dipole model of {\sc Ariadne 4.12}~\cite{Sjostrand:1985ys} which uses the Lund string model of {\sc Jetset 7.4}~\cite{Sjostrand:1993yb} for the hadronisation. 

In order to improve the Monte Carlo description 
of the photoproduction component, the contribution from the direct
 subprocesses was enlarged from 3\% (default) to 9\% in the inclusive {\sc Pythia} sample
while contributions by diffractive subprocesses were reduced accordingly.
This procedure ensured that previous ZEUS results were reproduced~\cite{Derrick:1992kk,Chekanov:2001gw} and the predicted inclusive {\sc Pythia} cross section remained unchanged.
The predicted photoproduction cross sections for HER, MER and LER were then 
validated against photoproduction data samples selected using the 6m-tagger.  
The predicted cross sections were consistent with these data within 
the $\pm10\%$ total uncertainty on the data.  

\begin{figure}
\begin{center}
\includegraphics[angle=0,scale=0.75]{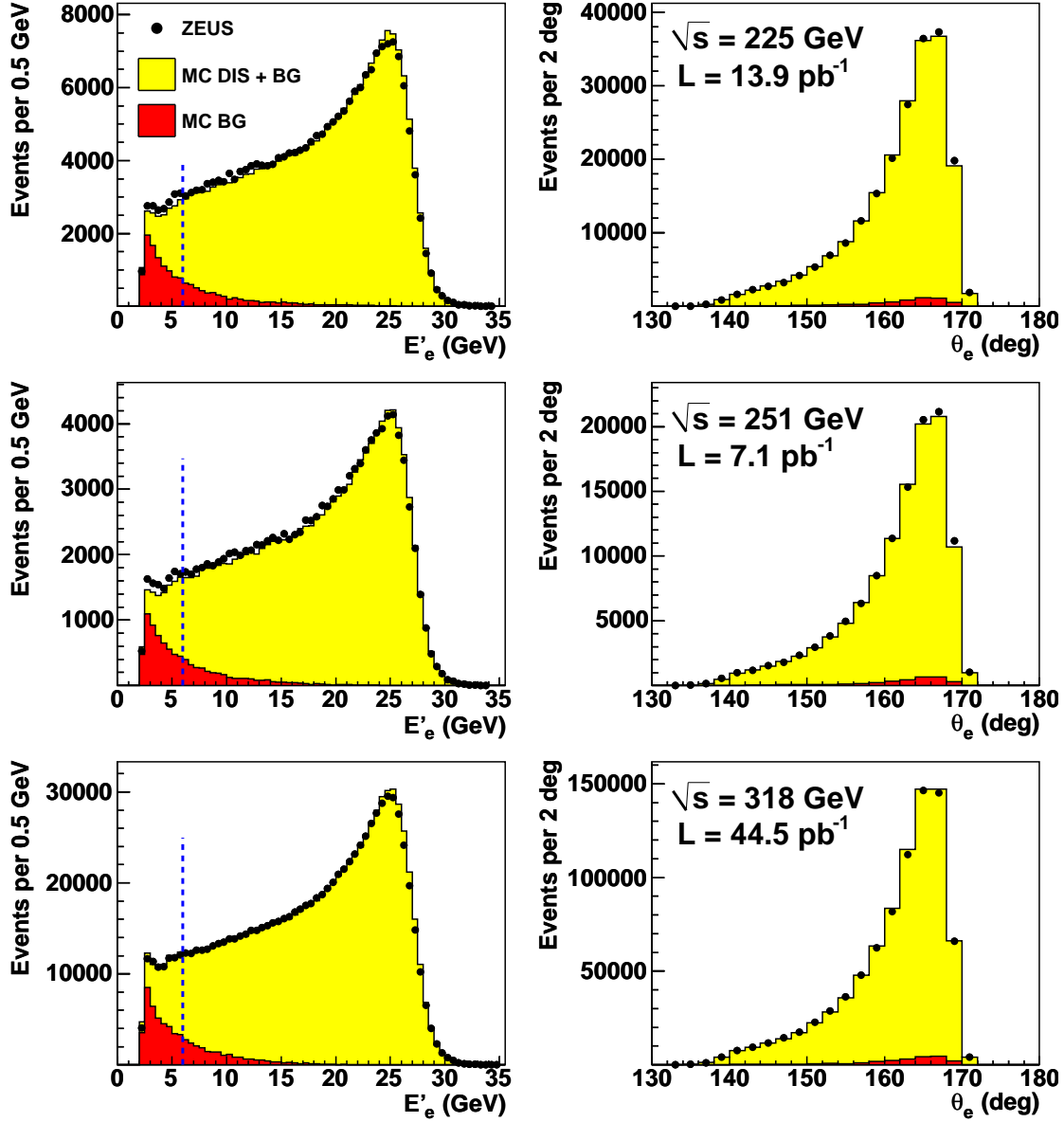}
\caption{Detector-level distributions of the energy, $E^{\prime}_e$, and polar angle, $\theta_e$, of the scattered electron candidates 
within the HER, MER and LER data sets compared to the combined MC predictions (MC DIS+BG).
The background only MC is labelled MC BG. 
The vertical dashed-line represents the $E^{\prime}_e$ cut. The $\theta_e$ 
distributions are shown for $E^\prime_e \geq 6$\,GeV.
  \label{fig:control}}
\end{center}
\end{figure}

Figure~\ref{fig:control} shows the distributions of the variables $E^{\prime}_e$ 
and $\theta_e$ within the HER, MER and LER data sets compared to the combined 
detector-level predictions from the MC models. The agreement is good in 
all cases. According to the MC models, the final data sample 
contained 97\% DIS signal and 3\% background events.  
The vast majority of the background events were found at low $Q^2$ and 
high $y$; in the most affected kinematic bin, the background fraction 
was 16\%.

\begin{figure}
\begin{center}
\includegraphics[angle=0,scale=0.75]{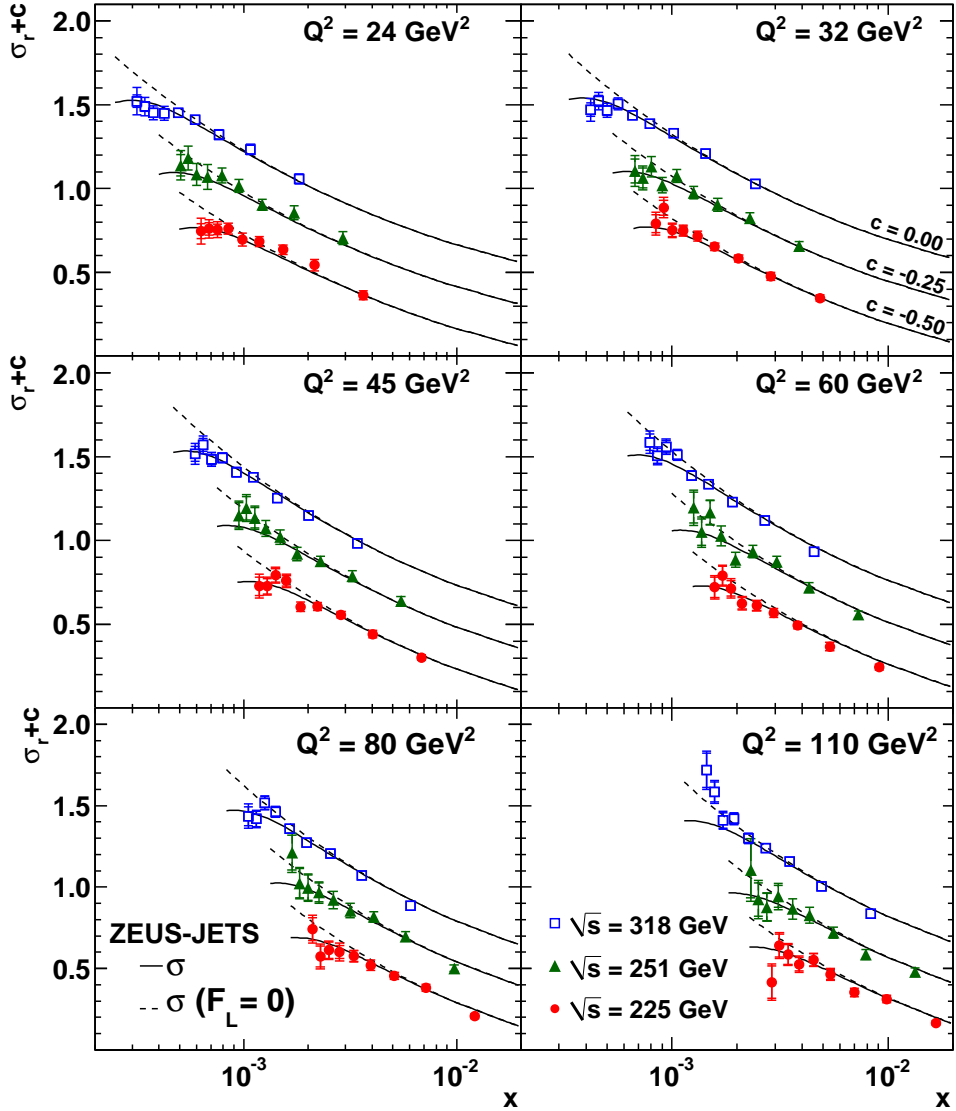}
\caption{The reduced cross sections at 6 values of $Q^2$ as a function 
of $x$ for the three 
centre-of-mass energies.
The points represent the ZEUS data from 
the HER ($\square$), MER ($\blacktriangle$) and LER ($\bullet$), respectively.
The solid lines represent the predicted reduced cross sections, using the 
ZEUS-JETS PDFs. The dashed lines represent the predicted reduced cross sections 
when $F_L$ is set to zero.   
The points and lines are shifted by $c$ (see top right) for 
clarity. The inner error bars represent the statistical uncertainty.  
The outer error bars represent the statistical plus systematic uncertainties 
added in quadrature. A further $\pm2.7$\% systematic normalisation uncertainty 
is not included. 
  \label{fig:reduced_Xsecs}}
\end{center}
\end{figure}

The reduced cross sections, $\tilde{\sigma}$, were measured from 
the HER, MER and LER samples in the kinematic region $0.09<y<0.78$ 
and $20<Q^2<130$\gev$^2$.  The $\tilde{\sigma}$ are given double 
differentially in $x$ and $Q^2$ in Tables~\ref{tab:sigHER}--\ref{tab:sigLER}. 
The $\tilde{\sigma}$ are also shown at the $6$ selected $Q^2$ values as 
functions of $x$ in Fig.~\ref{fig:reduced_Xsecs}.
The cross sections have been compared to 
DGLAP-predictions based on 
the NLO ($\mathcal{O}(\alpha_s^2)$) ZEUS-JETS PDF set~\cite{Chekanov:2005nn}, as well as the prediction 
for $F_L\equiv 0$. The QCD prediction with a non-zero $F_L$, 
describes
the data well and 
is favoured over
$F_L\equiv 0$.  

\section{Systematic uncertainties}
\label{sec:syst}

The systematic uncertainty on the reduced cross sections due to the following 
sources 
were evaluated~\cite{thesis:shima} (the numbers in the 
parentheses are the maximum uncertainty observed in any one of the reduced 
cross section bins):  

\begin{itemize}
\item $\{\delta_{\gamma p}\}$, the $\pm10\%$ uncertainty on the level of photoproduction background ($-2\%$);
\item $\{\delta_{E_e}\}$, the electron energy-scale uncertainty 
of $\pm 0.5\%$ for $E^{\prime}_e>20$\,GeV, increasing to $\pm 1.9\%$ 
at $E^{\prime}_e=6$\gev ($4.4\%$);
\item $\{\delta_{E_{h}}\}$, the $\pm2\%$ hadronic energy-scale uncertainty ($-4.1\%$);
\item $\{\delta_{eID}\}$, the uncertainty on the electron-finding efficiency, 
  evaluated by loosening (tightening) the criterion applied to the output 
of the neural network used to select electron candidates, 
both
for 
data and MC ($\pm 1.8\%$);
\item $\{\delta_{dx},~\delta_{dy}\}$, the SRTD and HES position uncertainty 
of $\pm 2$~mm in both the horizontal and vertical directions ($\pm3\%$);
\item $\{\delta_{\rm MVD},~\delta_{\rm CTD}\}$, the  uncertainty on the hit-finding 
efficiency, evaluated by  loosening (tightening) the hit fraction criteria, 
both
for 
data and MC ($+3.7\%$);
\item $\{\delta_{\rm diff}\}$, the $\pm10\%$ uncertainty on  the scale factors applied to the diffractive {\sc Djangoh} component ($-0.7\%$).
\end{itemize}

The one-standard-deviation systematic uncertainties due to each source 
are listed in \mbox{Tables~\ref{tab:sigHER}}--\ref{tab:sigLER} for the 
reduced cross sections at the three different centre-of-mass energies.
All of the uncertainties are symmetric. 
They are quoted with a sign indicating how the reduced cross sections would vary given an upwards variation in the electron or hadronic energy scales, the SRTD and HES positions, the photoproduction cross section or the diffractive scale factors, or looser selection criteria on the neural network output or MVD or CTD hit fractions.

The total uncertainty on the normalisation included
\begin{itemize}
\item the luminosity uncertainty, which was $\pm2.6\%$ for all three data sets, of which $\pm1\%$ was uncorrelated between the data sets; 
\item the uncertainty on simulating the interaction-vertex distribution, 
evaluated by comparing the ratio of the number of events 
with $|Z_{\rm vtx}|\le30$~cm and $|Z_{\rm vtx}|>30$~cm in data 
and MC ($\pm0.3\%$);
\item the trigger-efficiency uncertainty ($\pm0.5\%$).
\end{itemize}
The luminosity, vertex-distribution and trigger-efficiency uncertainties 
are perfectly correlated between bins and hence, when 
added in quadrature, 
constitute 
a total normalisation uncertainty of $\pm2.7$\%, of which $\pm2.5$\% was 
correlated between the running periods and $\pm1.1$\% uncorrelated.
The uncertainty due the electroweak corrections was found to be negligible.



The total systematic uncertainty in each bin, 
formed by adding the individual uncertainties in quadrature,
is also given in Tables~\ref{tab:sigHER}--\ref{tab:sigLER}. 
This sum also includes the statistical uncertainty due to the combined MC 
sample $\{\delta_{unc}\}$ and is the only systematic source that is considered to be 
uncorrelated between bins. 
This total systematic uncertainty 
does not include the total normalisation uncertainty.  
Propagation of the systematic uncertainties to $F_L$, $F_2$ and $R$ is described in the next section.


\section{\boldmath Extraction of $F_L$, $F_2$ and $R$}

In order to extract $F_L$, $F_2$ and $R$ a different binning scheme
than that given in Tables~\ref{tab:sigHER}--\ref{tab:sigLER} was applied 
to the reduced cross sections. Bins  in $y$  were chosen such that, for 
each of the $6$ $Q^2$ bins, there were $3$ values of $x$ at which the 
reduced cross sections were measured from all three data sets.  
This removed the need to interpolate the data between different points 
in the $(x,Q^2)$ plane. The structure functions were extracted by 
performing a simultaneous fit 
to
these $54$ measured cross section 
values using Eq.~\ref{eq:disnc-xsec}. Prior to fitting, the three data 
sets were normalised to their luminosity-weighted average in the restricted 
kinematic region, $y<0.3$, where the contribution to the reduced 
cross sections from $F_L$ is small. This procedure resulted in scaling 
the data by factors of $1.0027\pm 0.0027$, $0.9869\pm 0.0051$ 
and $0.9997\pm 0.0039$, for the HER, MER and LER data sets, respectively.  
The spread of these factors is consistent with the uncorrelated part of 
the total normalisation uncertainty
of 1.1\%.
To extract $F_L$ and $F_2$, $48$ parameters were fit simultaneously: 
$18$ $F_2$ and $18$ $F_L$ values for the $18$ $(x,Q^2)$ points;
$3$ relative normalisation factors for the HER, MER and LER data sets 
and $9$ global shifts of systematic uncertainties ($\delta_{\gamma p}$, $\delta_{E_e}$,
$\delta_{E_{h}}$, $\delta_{eID}$, $\delta_{dx}$, $\delta_{dy}$,
$\delta_{\rm MVD}$, $\delta_{\rm CTD}$, $\delta_{\rm diff}$). The three
normalisation factors allowed for variations of the relative
normalisation factors within their remaining uncertainties (see above).
The nine global shifts allowed for changes in the central values of
$\tilde{\sigma}$ in a correlated manner across the $(x,Q^2)$ plane according to the
uncertainties listed.  
The probability
distributions for the shifts of the systematic sources and the
relative normalisations were taken to be Gaussian, with standard
deviations equal to the corresponding systematic uncertainty.  The
probability distributions for the cross sections at each $(x,Q^2)$
point were also taken to be Gaussian with standard deviations given by
$\delta_{stat}$ and $\delta_{unc}$ added in quadrature. 
The fit
was performed within the BAT (Bayesian Analysis Toolkit) package~\cite{Caldwell:2008fw} which, using
a Markov chain MC, scans the full posterior probability density function 
in the $48$-dimensional parameter space.

\begin{figure}
\begin{center}
\includegraphics[angle=0,scale=0.75]{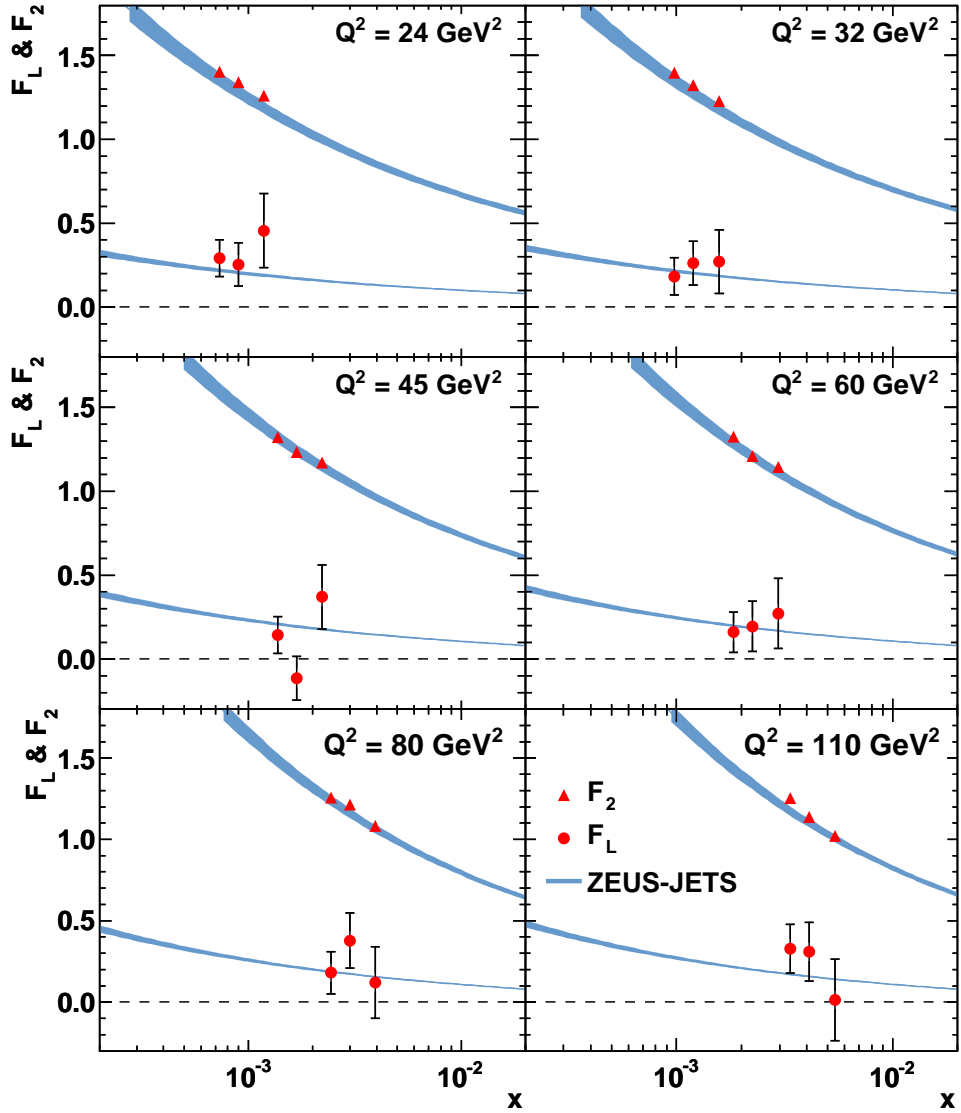}
\caption{$F_L$ and $F_2$ at 6 values of $Q^2$ as a function of $x$. 
The points represent the ZEUS data for $F_L$ ($\bullet$) and $F_2$ 
($\blacktriangle$), respectively.
The error bars on the data represent the combined statistical and 
systematic uncertainties. The error bars on $F_2$ are smaller than the
symbols. A further $\pm2.5$\% correlated normalisation 
uncertainty is not included. 
%
The DGLAP-predictions for $F_L$ and $F_2$ using the ZEUS-JETS PDFs are also shown. 
The bands indicate the uncertainty in the predictions.
  \label{fig:F2FL_x}}
\end{center}
\end{figure}

Initially, the $F_L$ and $F_2$ parameters were left unconstrained 
and flat prior probabilities were assumed. The results are given 
in Table~\ref{tab:FL_F2}, and are labelled with the superscript~$(1)$. 
The values quoted in the table were evaluated at the 
point where the probability density function attains its global maximum.
The uncertainty ranges correspond 
to minimal 68\% probability intervals.
These ranges represent the full experimental uncertainty, which comprises 
statistical as well as systematic uncertainties. 
%
The fitted shifts, representing the correlated variation of the 
data points according to relative normalisation and correlated 
systematic uncertainties, 
are typically within 0.1 and at most 0.5 standard deviations of the
normalisation or systematic uncertainties. 
The $F_2$ values typically 
have uncertainties of $0.03$, while the $F_L$ values have uncertainties
ranging from $0.1$ to $0.2$. These $F_2$ measurements are the most
precise available from the ZEUS collaboration in the kinematic region 
studied here. The results are shown in Fig.~\ref{fig:F2FL_x} 
together with predictions from the ZEUS-JETS PDF fit. 
Good agreement is observed.

Applying constrained priors $F_2 \ge 0$ and $0 \; \leq \; F_L \; \leq \; F_2$ in
the fitting gave marginally 
different results as seen in
Table~\ref{tab:FL_F2} (results are denoted with the superscript~$(2)$).
For example, the most probable value for $F_L$ at 
$Q^2=45$\gev$^2$ and $x=0.00153$ is now $0$, in which case, 
a 68\% probability upper limit is given. 

\begin{figure}
\begin{center}
\includegraphics*[angle=0,scale=0.75]{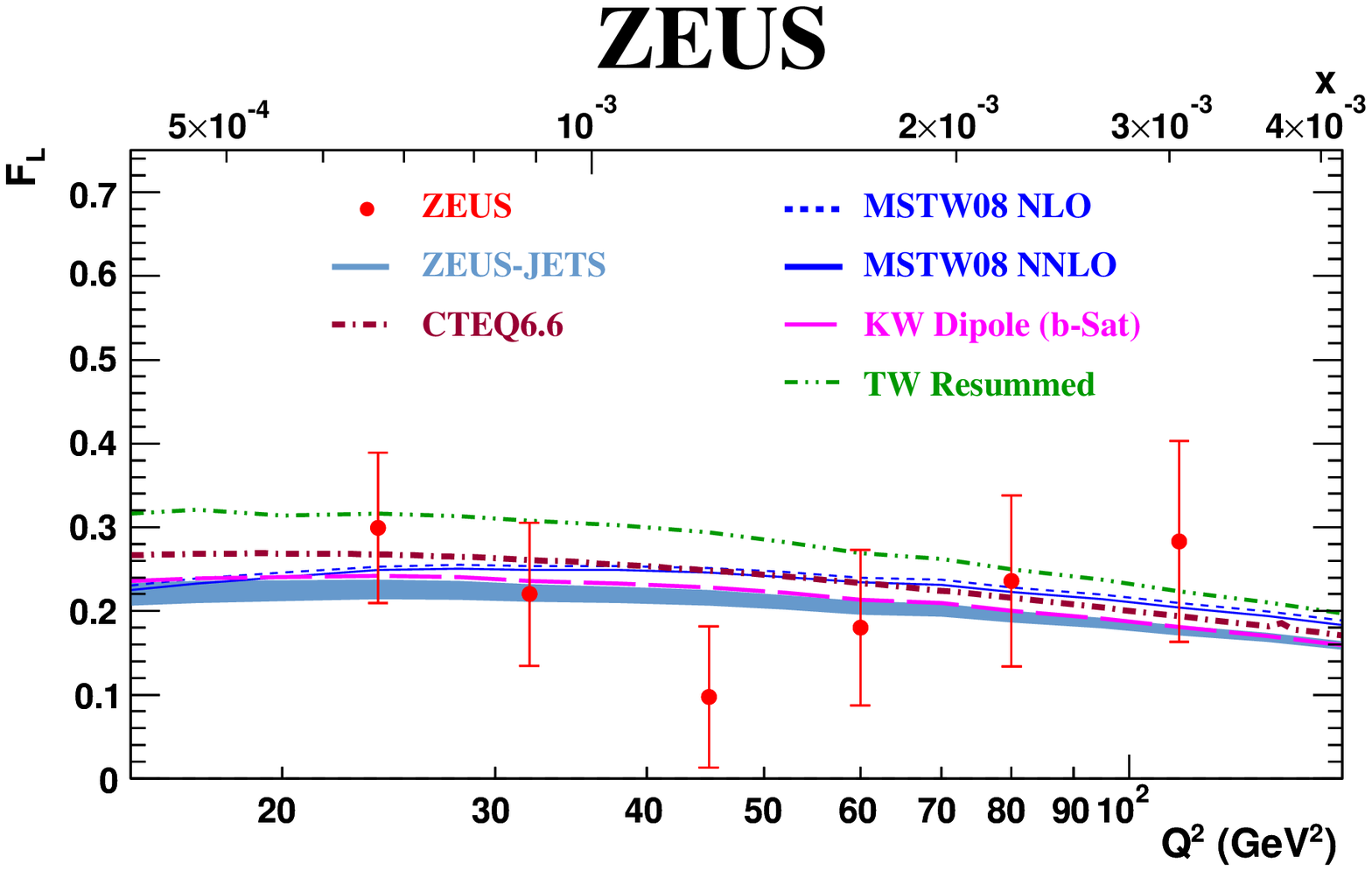}
\put(-40,210){(a)}%
\vspace{-0.7cm}
\includegraphics*[angle=0,scale=0.75]{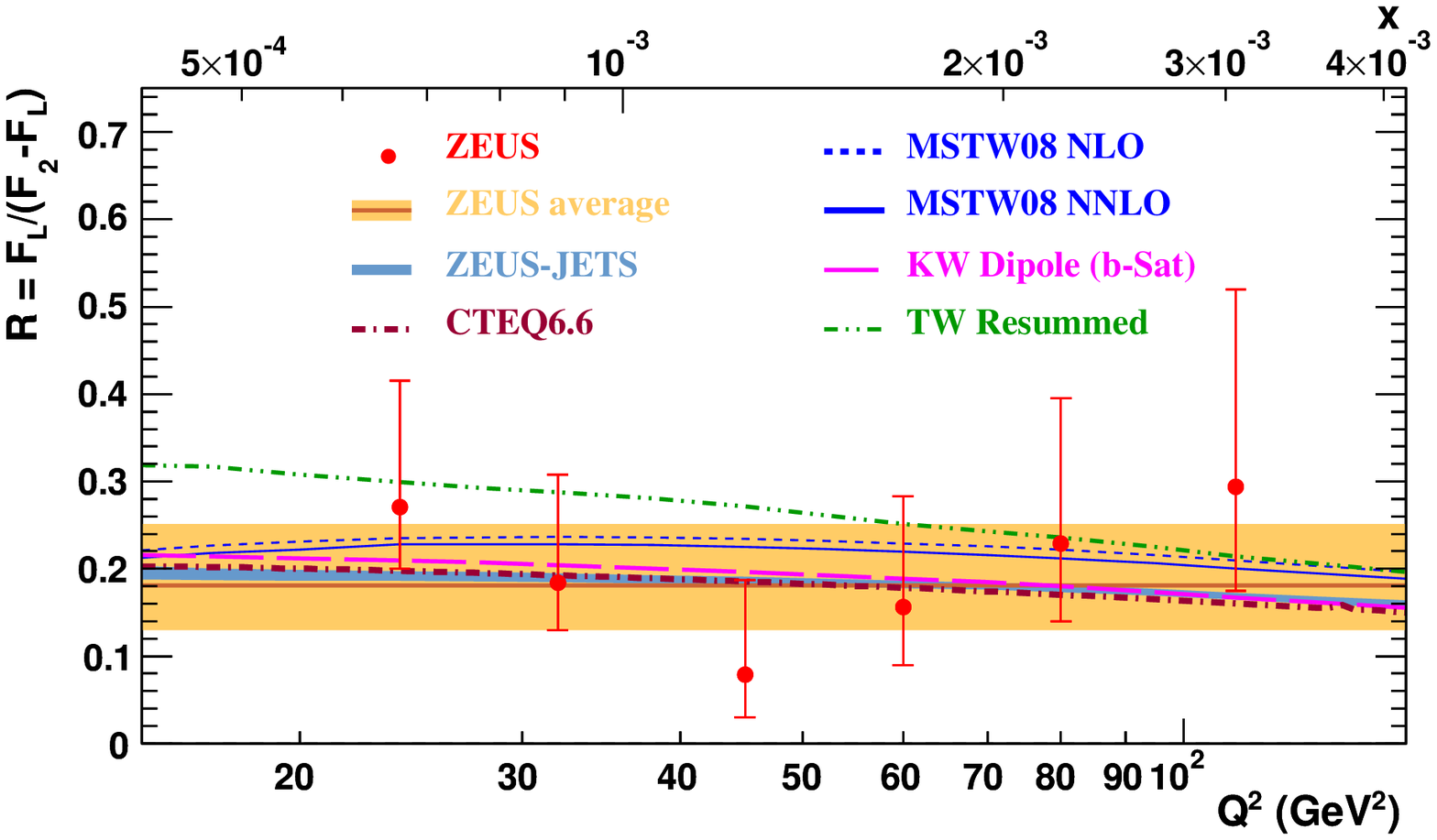}
\put(-40,210){(b)}%
\caption{
Values of (a) $F_L$ and (b) $R$ as a function of $Q^2$. 
The error bars on the data represent the combined statistical 
and systematic uncertainties. 
A further $\pm2.5$\%  correlated normalisation uncertainty is not 
included.
The shaded band labelled ZEUS average 
represents the 68\% probability interval for the overall~$R$.
The lines represent various model predictions (see text for details).
  \label{fig:F2FL_R_Q2}}
\end{center}
\end{figure}

Further fits to the data were performed to extract $F_L(Q^2)$, 
$R(Q^2)$, 
and a single overall value of $R$ for the full data set. In each case, the 
same fitting procedure as described above was used, but with a reduced number 
of parameters. To extract $F_L(Q^2)$, first $r(Q^2)$ was fitted, 
where $r=F_L/F_2$. In fitting $r(Q^2)$, a single value of $r$ was 
taken for all $x$ points in the same $Q^2$ bin. 
Only a weak dependence of $r$ on $x$ in a restricted $x$ range
is expected in the NLO DGLAP formalism as well as in phenomenological
models. This prediction is supported by the data within the
experimental uncertainties.
Flat prior distributions 
for $r(Q^2)$ were assumed and both unconstrained and constrained fits were 
made, with $r(Q^2)\ge 0$ enforced in the latter. The value of $F_L(Q^2)$ was 
then evaluated as $F_L(x_i,Q^2)=r(Q^2)F_2(x_i,Q^2)$, where for each $Q^2$ point, 
$x_i$ was chosen such that $Q^2/x_i$ was constant, which for $\sqrt{s}=225$\gev, corresponds to $y=0.71$. 
The results are given in Table~\ref{tab:FL_Q2} and the unconstrained values 
are shown in Fig.~\ref{fig:F2FL_R_Q2}a. These data are in good agreement with 
the results obtained by the H1 collaboration~\cite{Aaron:2008tx}.

Values of $R(Q^2)$ and an overall value of $R$ were extracted with 
flat prior distributions. Both unconstrained and constrained fits were made. 
In the latter, it was required that $0\le F_L(Q^2)\le F_2(Q^2)$ and $0\le F_L\le F_2$. The results 
from both fits are given in Table~\ref{tab:R_Q2} and the unconstrained 
$R(Q^2)$ values are shown in Fig.~\ref{fig:F2FL_R_Q2}b.
The uncertainty in the overall $R$ is not reduced as much as might be
expected compared to the uncertainties on $R(Q^2)$ due to the
correlation between the values at different $Q^2$.
The value of $R$ from both the unconstrained and constrained fits was 
$R=0.18^{+0.07}_{-0.05}$. 

Figures~\ref{fig:F2FL_R_Q2}a and \ref{fig:F2FL_R_Q2}b also show a comparison 
of the data with predictions based on the ZEUS-JETS and 
CTEQ6.6~\cite{Nadolsky:2008zw} NLO and MSTW08~\cite{Martin:2009iq} NLO and 
NNLO\footnote{Based on the NNLO calculations by Moch, Vermaseren and 
Vogt~\cite{Moch:2004xu,Vermaseren:2005qc}.} fits. All these predictions 
are based on the DGLAP formalism\footnote{The conventions used for the CTEQ6.6, ZEUS-JETS and MSTW08 NLO curves are not the same, for example,  $F_L$ in  CTEQ6.6 is  calculated to $\mathcal{O}(\alpha_s)$ whereas  $F_L$ in the ZEUS-JETS and MSTW08 fits are  calculated to $\mathcal{O}(\alpha_s^2)$ .   This  accounts for most of the differences.}. 
Also shown are predictions from the NLL BFKL resummation fit from 
Thorne and White~(TW)~\cite{White:2006yh}, and
the prediction from the impact-parameter-dependent 
dipole saturation model (b-Sat) of Kowalski and Watt based on DGLAP evolution of the gluon density~\cite{Watt:2007nr,*Kowalski:2006hc,*Kowalski:2003hm}. 
%
All of the models are consistent with the data. 


\section{Summary}


The first measurement of $F_L(x,Q^2)$ by the ZEUS collaboration 
is presented, as is the first measurement of $F_2(x,Q^2)$ at low $x$ 
that does not include assumptions about $F_L$.  
The $F_2$ values are the most precise available from the ZEUS collaboration 
in the kinematic region studied.
The extraction of $F_L$ and $F_2$ was based on the reduced 
double-differential cross sections, $\tilde{\sigma}(x,Q^2)$, which were measured 
for $0.09<y<0.78$ and $20<Q^2<130$\gev$^2$ using 
data collected at $\sqrt{s}=318$, 251 and 225\gev. 
In addition, $F_L$ and the ratio, $R$, have been extracted 
as a function of $Q^2$. An overall value of $R=0.18^{+0.07}_{-0.05}$ was 
extracted for the entire kinematic region studied. 
A wide range of theoretical predictions agree with the measured $F_L$.
The measurements provide strong evidence of a non-zero $F_L$.

%
%

%
%
\section*{Acknowledgements}

We appreciate the contributions to the construction and maintenance
of the ZEUS detector of many people who are not listed as authors.
The HERA machine group and the DESY computing staff
are especially acknowledged for their success
in providing excellent operation of the collider
and the data-analysis environment.
We thank the DESY directorate for their strong support and encouragement.


{
\def\bibname{\Large\bf References}
\def\refname{\Large\bf References}
\pagestyle{plain}
\bibliographystyle{./l4z_default}
{\raggedright
\bibliography{./syn.bib,%
              ../BiBTeX/user/syn.bib,%
              ../BiBTeX/bib/l4z_articles.bib,%
              ../BiBTeX/bib/l4z_books.bib,%
              ../BiBTeX/bib/l4z_conferences.bib,%
              ../BiBTeX/bib/l4z_h1.bib,%
              ../BiBTeX/bib/l4z_misc.bib,%
              ../BiBTeX/bib/l4z_old.bib,%
              ../BiBTeX/bib/l4z_preprints.bib,%
              ../BiBTeX/bib/l4z_replaced.bib,%
              ../BiBTeX/bib/l4z_temporary.bib,%
              ../BiBTeX/bib/l4z_zeus.bib}}
}
\vfill\eject


\begin{center}
\scriptsize
\begin{longtable}{|r|c|c||c|r|r||r|r|r|r|r|r|r|r|r|r|}
\hline
$Q^2$ & $x$ & $y$ & $\tilde{\sigma}^{e^+p}$ & $\delta_{stat}$ &$\delta_{sys}$ & $\delta_{unc}$ & $\delta_{\gamma p}$ & $\delta_{E_e}$ &$\delta_{E_{h}}$ &$\delta_{eID}$ & $\delta_{dx}$ & $\delta_{dy}$ & $\delta_{\rm MVD}$ &$\delta_{\rm CTD}$ &$\delta_{\rm diff}$\\
(GeV$^2$) & & & HER & (\%) & (\%) & (\%) & (\%) & (\%) & (\%) & (\%) & (\%) & (\%) & (\%) & (\%) & (\%) \\\hline
  24 & $1.82\cdot 10^{-3}$ & 0.13 & 1.057 &  1.0 &  2.5 &  0.9 &  0.0 &  1.0 & -0.8 &  0.0 &  0.8 &  0.3 & -0.4 &  1.7 & -0.3 \\
  24 & $1.08\cdot 10^{-3}$ & 0.22 & 1.234 &  0.8 &  2.4 &  0.8 & 0.0 &  1.6 & -0.2 & 0.0 &  0.2 &  0.3 & -0.6 &  1.4 & -0.3 \\
  24 & $7.63\cdot 10^{-4}$ & 0.31 & 1.321 &  0.8 &  2.2 &  0.7 & 0.0 & -0.2 & -0.2 & 0.0 & -0.3 &  0.1 & -1.8 &  1.0 & -0.4 \\
  24 & $5.92\cdot 10^{-4}$ & 0.40 & 1.410 &  0.8 &  1.6 &  0.7 & -0.1 & -0.4 & -0.3 &  0.3 & -0.7 &  0.1 & -0.3 &  1.0 & -0.4 \\
  24 & $4.93\cdot 10^{-4}$ & 0.48 & 1.453 &  0.8 &  1.5 &  0.8 & -0.3 &  0.3 & -0.5 & -0.1 & -0.2 &  0.1 &  0.3 &  0.9 & -0.4 \\
  24 & $4.23\cdot 10^{-4}$ & 0.56 & 1.448 &  0.9 &  2.7 &  1.0 & -0.6 &  0.4 & -0.8 &  0.6 & -0.4 &  0.2 &  1.0 &  1.9 & -0.5 \\
  24 & $3.76\cdot 10^{-4}$ & 0.63 & 1.452 &  1.1 &  2.7 &  1.2 & -0.8 &  0.1 & -1.4 & -0.6 &  0.3 & -0.1 &  0.7 &  1.4 & -0.5 \\
  24 & $3.43\cdot 10^{-4}$ & 0.69 & 1.489 &  1.3 &  3.6 &  1.5 & -1.2 &  0.5 & -2.4 & -0.6 & -0.3 & -0.1 &  0.7 &  1.2 & -0.6 \\
  24 & $3.16\cdot 10^{-4}$ & 0.75 & 1.521 &  1.5 &  5.1 &  2.0 & -2.0 &  0.4 & -3.9 & -0.7 &  0.5 & -0.4 &  1.3 & -0.5 & -0.7 \\
\hline
  32 & $2.43\cdot 10^{-3}$ & 0.13 & 1.027 &  0.6 &  1.8 &  0.6 &  0.0 &  1.1 & -0.6 &  0.0 & -0.5 &  0.1 & -0.8 &  0.7 & -0.2 \\
  32 & $1.43\cdot 10^{-3}$ & 0.22 & 1.209 &  0.6 &  2.0 &  0.6 & 0.0 &  1.5 & -0.2 &  0.0 & -0.6 &  0.1 & -0.5 &  0.8 & -0.3 \\
  32 & $1.02\cdot 10^{-3}$ & 0.31 & 1.331 &  0.7 &  1.6 &  0.6 & 0.0 & -0.1 & -0.2 &  0.1 & -0.1 & -0.1 &  0.4 &  1.3 & -0.4 \\
  32 & $7.89\cdot 10^{-4}$ & 0.40 & 1.388 &  0.8 &  1.5 &  0.7 & -0.1 & -0.6 & -0.2 &  0.2 &  0.4 & -0.1 &  0.1 &  1.0 & -0.4 \\
  32 & $6.57\cdot 10^{-4}$ & 0.48 & 1.435 &  0.9 &  1.7 &  0.8 & -0.2 & 0.0 & -0.3 & -0.5 & -0.5 & -0.1 & -0.8 &  0.8 & -0.4 \\
  32 & $5.63\cdot 10^{-4}$ & 0.56 & 1.504 &  1.0 &  2.2 &  1.0 & -0.4 & -0.4 & -0.7 &  0.1 & -0.1 & -0.2 & -1.5 &  0.6 & -0.5 \\
  32 & $5.01\cdot 10^{-4}$ & 0.63 & 1.465 &  1.3 &  2.6 &  1.5 & -0.7 &  1.2 & -1.3 &  0.2 & -0.1 &  0.1 & -0.6 & -0.4 & -0.5 \\
  32 & $4.57\cdot 10^{-4}$ & 0.69 & 1.522 &  1.4 &  3.1 &  1.6 & -0.8 & -0.9 & -1.8 &  0.3 & -0.3 & -0.2 & -1.3 & -0.4 & -0.6 \\
  32 & $4.21\cdot 10^{-4}$ & 0.75 & 1.470 &  1.7 &  4.4 &  2.2 & -1.7 &  1.5 & -2.5 & -0.9 & -0.4 &  0.3 & -1.0 & -0.7 & -0.5 \\
\hline
  45 & $3.41\cdot 10^{-3}$ & 0.13 & 0.984 &  0.6 &  1.6 &  0.5 &  0.0 &  0.9 & -0.7 &  0.0 & -0.1 &  0.1 &  0.1 &  1.0 & -0.2 \\
  45 & $2.02\cdot 10^{-3}$ & 0.22 & 1.151 &  0.6 &  2.0 &  0.5 &  0.0 &  1.5 & -0.2 &  0.0 & -0.2 & -0.2 & -0.6 &  0.9 & -0.3 \\
  45 & $1.43\cdot 10^{-3}$ & 0.31 & 1.253 &  0.7 &  1.4 &  0.6 & 0.0 & -0.2 & -0.1 &  0.0 & -0.3 &  0.1 & -0.7 &  0.8 & -0.4 \\
  45 & $1.11\cdot 10^{-3}$ & 0.40 & 1.376 &  0.9 &  1.3 &  0.8 & -0.1 & -0.5 & -0.2 & -0.1 & -0.4 &  0.2 &  0.1 &  0.7 & -0.4 \\
  45 & $9.24\cdot 10^{-4}$ & 0.48 & 1.408 &  1.0 &  1.8 &  0.9 & -0.1 & -1.0 & -0.3 & -0.2 &  0.4 &  0.0 &  0.5 &  0.9 & -0.5 \\
  45 & $7.92\cdot 10^{-4}$ & 0.56 & 1.492 &  1.1 &  1.8 &  1.1 & -0.3 & -0.2 & -0.4 & -0.4 &  0.5 &  0.1 &  0.6 &  0.8 & -0.4 \\
  45 & $7.04\cdot 10^{-4}$ & 0.63 & 1.483 &  1.4 &  2.5 &  1.5 & -0.6 &  0.4 & -1.0 & -1.1 &  0.5 & -0.3 &  0.5 &  0.3 & -0.5 \\
  45 & $6.43\cdot 10^{-4}$ & 0.69 & 1.571 &  1.6 &  3.1 &  1.9 & -0.8 & -0.3 & -1.3 &  0.6 &  0.5 &  0.1 &  1.6 & -0.4 & -0.5 \\
  45 & $5.92\cdot 10^{-4}$ & 0.75 & 1.517 &  1.8 &  3.7 &  2.3 & -1.4 &  0.9 & -1.8 &  0.4 & -0.2 &  0.2 & -0.7 & -1.2 & -0.6 \\
\hline
  60 & $4.55\cdot 10^{-3}$ & 0.13 & 0.932 &  0.7 &  1.7 &  0.6 &  0.0 &  0.9 & -0.6 &  0.0 & -0.5 & -0.3 & -0.6 &  0.7 & -0.1 \\
  60 & $2.69\cdot 10^{-3}$ & 0.22 & 1.119 &  0.7 &  1.9 &  0.6 &  0.0 &  1.5 & -0.1 &  0.0 & -0.4 & 0.0 & -0.7 &  0.5 & -0.3 \\
  60 & $1.91\cdot 10^{-3}$ & 0.31 & 1.231 &  0.9 &  1.1 &  0.7 & 0.0 &  0.2 & -0.1 & 0.0 & -0.2 &  0.2 &  0.3 &  0.6 & -0.3 \\
  60 & $1.48\cdot 10^{-3}$ & 0.40 & 1.337 &  1.0 &  1.6 &  0.9 & 0.0 & -1.1 & -0.2 &  0.1 & -0.6 & -0.3 &  0.3 &  0.4 & -0.4 \\
  60 & $1.23\cdot 10^{-3}$ & 0.48 & 1.388 &  1.2 &  1.7 &  1.1 & -0.1 & -0.5 & -0.2 &  0.3 &  0.2 &  0.3 & -0.8 & -0.6 & -0.5 \\
  60 & $1.06\cdot 10^{-3}$ & 0.56 & 1.510 &  1.3 &  1.8 &  1.2 & -0.2 & -0.7 & -0.5 & 0.0 &  0.3 &  0.2 & -0.8 &  0.1 & -0.4 \\
  60 & $9.39\cdot 10^{-4}$ & 0.63 & 1.560 &  1.7 &  2.6 &  1.7 & -0.4 &  0.6 & -0.7 &  0.6 & -0.6 &  0.3 & -0.9 & -1.0 & -0.5 \\
  60 & $8.57\cdot 10^{-4}$ & 0.69 & 1.504 &  1.9 &  3.1 &  2.2 & -0.9 &  1.0 & -1.3 & -0.2 &  0.1 &  0.4 & -0.5 & -0.4 & -0.5 \\
  60 & $7.89\cdot 10^{-4}$ & 0.75 & 1.586 &  2.1 &  3.6 &  2.2 & -0.9 & -1.7 & -1.6 & -0.6 & -0.4 & -0.3 & -0.5 & -0.9 & -0.6 \\
\hline
\multicolumn{16}{r}{{Continued on Next Page\ldots}} \\
\newpage
\multicolumn{16}{l}{{\ldots Continued from Previous Page}} \\
\hline
$Q^2$ & $x$ & $y$ & $\tilde{\sigma}^{e^+p}$ & $\delta_{stat}$ &$\delta_{sys}$ & $\delta_{unc}$ & $\delta_{\gamma p}$ & $\delta_{E_e}$ &$\delta_{E_{h}}$ &$\delta_{eID}$ & $\delta_{dx}$ & $\delta_{dy}$ & $\delta_{\rm MVD}$ &$\delta_{\rm CTD}$ &$\delta_{\rm diff}$\\
(GeV$^2$) & & & HER & (\%) & (\%) & (\%) & (\%) & (\%) & (\%) & (\%) & (\%) & (\%) & (\%) & (\%) & (\%) \\\hline
  80 & $6.07\cdot 10^{-3}$ & 0.13 & 0.884 &  0.8 &  1.6 &  0.6 &  0.0 &  0.9 & -0.7 &  0.0 &  0.1 &  0.3 & -0.7 &  0.5 & -0.1 \\
  80 & $3.59\cdot 10^{-3}$ & 0.22 & 1.071 &  0.8 &  1.7 &  0.7 &  0.0 &  1.5 & -0.2 &  0.0 & -0.3 & -0.1 & -0.1 &  0.1 & -0.2 \\
  80 & $2.54\cdot 10^{-3}$ & 0.31 & 1.204 &  1.0 &  1.2 &  0.8 &  0.0 & -0.2 & -0.1 & -0.3 & -0.1 & -0.1 & -0.6 & -0.2 & -0.3 \\
  80 & $1.97\cdot 10^{-3}$ & 0.40 & 1.273 &  1.2 &  1.2 &  1.0 & -0.1 & -0.4 & -0.1 & 0.0 &  0.0 &  0.1 &  0.2 &  0.1 & -0.4 \\
  80 & $1.64\cdot 10^{-3}$ & 0.48 & 1.358 &  1.4 &  1.6 &  1.2 & -0.1 & -0.5 & -0.3 &  0.1 & -0.1 & -0.3 &  0.4 &  0.5 & -0.4 \\
  80 & $1.41\cdot 10^{-3}$ & 0.56 & 1.463 &  1.5 &  1.9 &  1.4 & -0.2 &  0.5 & -0.3 &  0.2 & -0.6 & -0.2 &  0.3 &  0.5 & -0.5 \\
  80 & $1.25\cdot 10^{-3}$ & 0.63 & 1.517 &  1.9 &  2.1 &  1.8 & -0.1 & -0.6 & -0.5 & -0.3 &  0.4 & -0.1 & -0.3 & -0.2 & -0.5 \\
  80 & $1.14\cdot 10^{-3}$ & 0.69 & 1.419 &  2.2 &  3.2 &  2.5 & -0.8 &  0.1 & -0.8 &  0.3 & -0.1 & -0.5 &  0.4 &  1.3 & -0.4 \\
  80 & $1.05\cdot 10^{-3}$ & 0.75 & 1.436 &  2.5 &  4.6 &  3.2 & -1.2 &  1.9 & -1.9 &  0.6 & -0.4 & -0.2 &  0.9 & -0.9 & -0.5 \\
\hline
 110 & $8.34\cdot 10^{-3}$ & 0.13 & 0.837 &  0.9 &  1.3 &  0.8 &  0.0 &  0.7 & -0.7 &  0.0 & -0.2 &  0.2 & -0.1 &  0.2 & -0.1 \\
 110 & $4.93\cdot 10^{-3}$ & 0.22 & 1.005 &  1.0 &  1.7 &  0.9 &  0.0 &  1.4 & -0.2 & -0.1 & -0.2 & -0.2 &  0.1 &  0.1 & -0.2 \\
 110 & $3.50\cdot 10^{-3}$ & 0.31 & 1.157 &  1.2 &  1.5 &  1.0 &  0.0 & -0.5 & -0.1 &  0.1 &  0.3 &  0.2 & -0.7 & -0.6 & -0.3 \\
 110 & $2.71\cdot 10^{-3}$ & 0.40 & 1.240 &  1.4 &  1.5 &  1.2 & -0.1 & -0.7 &  0.1 &  0.0 & -0.3 &  0.1 &  0.2 &  0.2 & -0.4 \\
 110 & $2.26\cdot 10^{-3}$ & 0.48 & 1.300 &  1.6 &  1.9 &  1.5 & -0.2 & -0.4 &  0.0 & -0.1 & -0.4 &  0.3 & -0.8 & -0.3 & -0.4 \\
 110 & $1.94\cdot 10^{-3}$ & 0.56 & 1.420 &  1.8 &  2.0 &  1.7 & -0.2 &  0.6 & -0.2 &  0.1 & -0.7 & -0.2 & 0.0 & -0.5 & -0.4 \\
 110 & $1.72\cdot 10^{-3}$ & 0.63 & 1.409 &  2.3 &  3.2 &  2.4 & -0.4 & -1.6 & -0.3 &  0.5 &  0.5 &  0.1 &  0.3 & -0.7 & -0.4 \\
 110 & $1.57\cdot 10^{-3}$ & 0.69 & 1.584 &  2.6 &  3.5 &  2.8 & -0.7 & -0.8 & -0.6 &  0.7 & -0.2 & -0.3 &  0.4 &  1.4 & -0.4 \\
 110 & $1.45\cdot 10^{-3}$ & 0.75 & 1.717 &  4.1 &  5.5 &  4.6 & -0.9 &  1.5 &  0.9 &  1.2 &  0.8 & -0.1 & -0.4 & -1.7 & -0.4 \\
\hline
\multicolumn{16}{c}{\parbox{1.5\LTcapwidth}{}}\\
\multicolumn{16}{c}{\parbox{1.5\LTcapwidth}{
\normalsize{\bf Table 1: }\it
The reduced cross section, $\tilde{\sigma}$, for the 
reaction $e^+p\rightarrow e^+X$ at $\sqrt{s}=318$\gev.  
The first three columns contain the bin centres in $Q^2$, 
$x$ and $y$, the next three contain the measured cross section, 
the statistical uncertainty and the total systematic uncertainty, 
respectively.  The final ten columns list the uncorrelated, $\delta_{unc}$
and the bin-to-bin correlated uncertainties from each systematic source,  
$\delta_{\gamma p}$, $\delta_{E_e}$, $\delta_{E_{h}}$, $\delta_{eID}$, 
$\delta_{dx}$, $\delta_{dy}$, $\delta_{\rm MVD}$, $\delta_{\rm CTD}$, $\delta_{\rm diff}$.
For details, see Section~\ref{sec:syst}.
A further $\pm2.7$\% systematic normalisation uncertainty is not included, 
of which $\pm2.5$\% is correlated between the running periods and $\pm1.1$\% 
is uncorrelated.
}} 
\label{tab:sigHER}
\end{longtable}
\end{center}

\newpage

\begin{center}
\scriptsize
\begin{longtable}{|r|c|c||c|r|r||r|r|r|r|r|r|r|r|r|r|}
\hline
$Q^2$ & $x$ & $y$ & $\tilde{\sigma}^{e^+p}$ & $\delta_{stat}$ &$\delta_{sys}$ & $\delta_{unc}$ & $\delta_{\gamma p}$ & $\delta_{E_e}$ &$\delta_{E_{h}}$ &$\delta_{eID}$ & $\delta_{dx}$ & $\delta_{dy}$ & $\delta_{\rm MVD}$ &$\delta_{\rm CTD}$ &$\delta_{\rm diff}$\\
(GeV$^2$) & & & MER & (\%) & (\%) & (\%) & (\%) & (\%) & (\%) & (\%) & (\%) & (\%) & (\%) & (\%) & (\%) \\\hline
  24 & $2.91\cdot 10^{-3}$ & 0.13 & 0.955 &  2.4 &  3.3 &  1.3 &  0.0 &  0.9 & -0.7 & 0.0 &  1.1 &  0.5 & -1.6 &  2.0 & -0.2 \\
  24 & $1.72\cdot 10^{-3}$ & 0.22 & 1.105 &  2.1 &  3.0 &  1.1 & 0.0 &  1.6 & -0.3 &  0.2 &  1.4 & -0.3 & -1.4 &  1.1 & -0.3 \\
  24 & $1.22\cdot 10^{-3}$ & 0.31 & 1.153 &  2.1 &  2.1 &  1.0 & 0.0 & -0.6 & -0.2 &  0.0 & -0.3 & -0.6 & -0.6 &  1.4 & -0.3 \\
  24 & $9.47\cdot 10^{-4}$ & 0.40 & 1.267 &  2.1 &  2.3 &  1.0 & -0.1 &  0.1 & -0.2 &  0.5 &  1.1 &  0.2 &  0.8 &  1.3 & -0.4 \\
  24 & $7.89\cdot 10^{-4}$ & 0.48 & 1.328 &  2.2 &  2.6 &  1.2 & -0.3 & -0.7 & -0.4 & -0.4 & -0.3 & -0.2 &  0.7 &  1.9 & -0.4 \\
  24 & $6.76\cdot 10^{-4}$ & 0.56 & 1.319 &  2.4 &  5.0 &  1.3 & -0.5 &  0.7 & -0.8 &  0.2 & -0.5 &  0.4 &  2.8 &  3.7 & -0.4 \\
  24 & $6.01\cdot 10^{-4}$ & 0.63 & 1.334 &  3.0 &  3.8 &  1.6 & -0.7 & -0.3 & -1.4 & -0.9 &  1.2 &  0.6 &  2.0 &  1.6 & -0.5 \\
  24 & $5.49\cdot 10^{-4}$ & 0.69 & 1.432 &  3.2 &  3.9 &  1.8 & -1.0 & -0.5 & -2.0 & -0.2 & -0.3 &  0.6 &  0.8 &  2.2 & -0.6 \\
  24 & $5.05\cdot 10^{-4}$ & 0.75 & 1.389 &  3.9 &  5.0 &  2.5 & -2.0 & -0.8 & -3.5 &  0.9 &  0.2 &  0.4 & -0.4 &  0.8 & -0.6 \\
\hline
  32 & $3.88\cdot 10^{-3}$ & 0.13 & 0.909 &  1.7 &  2.1 &  0.8 &  0.0 &  1.1 & -0.8 &  0.0 & -0.2 & -0.1 & -0.3 &  1.3 & -0.2 \\
  32 & $2.29\cdot 10^{-3}$ & 0.22 & 1.074 &  1.7 &  2.4 &  0.8 & 0.0 &  1.6 & -0.3 &  0.1 & -0.8 &  0.2 & -0.6 &  1.1 & -0.2 \\
  32 & $1.63\cdot 10^{-3}$ & 0.31 & 1.153 &  1.9 &  2.9 &  0.9 & -0.1 & -0.3 & -0.1 & 0.0 &  0.8 &  0.9 &  1.0 &  2.3 & -0.4 \\
  32 & $1.26\cdot 10^{-3}$ & 0.40 & 1.226 &  2.1 &  2.2 &  1.0 & -0.1 & -0.3 & -0.2 &  0.4 &  0.1 &  0.2 &  0.6 &  1.6 & -0.4 \\
  32 & $1.05\cdot 10^{-3}$ & 0.48 & 1.326 &  2.3 &  1.9 &  1.2 & -0.2 & -0.6 & -0.3 & -0.1 &  0.2 &  0.2 & -0.2 &  1.2 & -0.4 \\
  32 & $9.01\cdot 10^{-4}$ & 0.56 & 1.270 &  2.7 &  2.5 &  1.4 & -0.5 &  0.6 & -0.8 & -0.8 & -0.8 & -0.9 & -0.5 &  0.7 & -0.4 \\
  32 & $8.01\cdot 10^{-4}$ & 0.63 & 1.381 &  3.4 &  2.9 &  1.9 & -0.5 &  0.6 & -1.0 &  0.6 & -1.6 & -0.2 &  0.4 &  0.5 & -0.6 \\
  32 & $7.32\cdot 10^{-4}$ & 0.69 & 1.314 &  3.9 &  3.7 &  2.2 & -0.9 &  0.5 & -1.7 &  0.4 &  0.1 & -0.8 & -1.7 &  1.0 & -0.5 \\
  32 & $6.73\cdot 10^{-4}$ & 0.75 & 1.355 &  4.5 &  5.0 &  2.7 & -1.6 &  1.6 & -2.8 & -1.1 &  0.9 & -0.3 & -1.6 &  0.3 & -0.6 \\
\hline
  45 & $5.46\cdot 10^{-3}$ & 0.13 & 0.890 &  1.5 &  2.4 &  0.7 &  0.0 &  1.2 & -0.7 & 0.0 & -0.3 &  0.3 &  0.3 &  1.7 & -0.1 \\
  45 & $3.23\cdot 10^{-3}$ & 0.22 & 1.037 &  1.7 &  2.5 &  0.8 & 0.0 &  1.2 & -0.2 & 0.0 & -0.4 &  0.2 &  0.9 &  1.8 & -0.2 \\
  45 & $2.29\cdot 10^{-3}$ & 0.31 & 1.126 &  1.9 &  1.8 &  0.9 & 0.0 &  0.2 & -0.1 & -0.2 & -1.0 & -0.5 & -0.5 &  0.8 & -0.3 \\
  45 & $1.77\cdot 10^{-3}$ & 0.40 & 1.171 &  2.3 &  2.5 &  1.1 & -0.1 & -0.4 &  0.1 &  0.3 & -1.5 & -0.1 &  0.6 &  1.4 & -0.4 \\
  45 & $1.48\cdot 10^{-3}$ & 0.48 & 1.270 &  2.7 &  2.3 &  1.3 & -0.2 & -0.5 & -0.3 &  0.3 &  0.7 &  0.5 &  0.4 &  1.4 & -0.5 \\
  45 & $1.27\cdot 10^{-3}$ & 0.56 & 1.323 &  3.0 &  2.1 &  1.5 & -0.3 & -0.5 & -0.6 &  0.1 &  0.4 & -0.8 & -0.3 &  0.6 & -0.5 \\
  45 & $1.13\cdot 10^{-3}$ & 0.63 & 1.385 &  3.8 &  3.3 &  2.2 & -0.6 & -0.8 & -1.1 &  1.8 &  0.4 &  0.5 &  0.2 &  0.3 & -0.4 \\
  45 & $1.03\cdot 10^{-3}$ & 0.69 & 1.443 &  4.2 &  3.7 &  2.4 & -0.8 & -0.6 & -1.3 &  0.2 &  1.1 &  0.4 &  1.2 &  1.6 & -0.6 \\
  45 & $9.47\cdot 10^{-4}$ & 0.75 & 1.400 &  4.7 &  3.8 &  2.6 & -1.0 &  0.7 & -1.9 & -1.0 &  0.6 &  0.8 &  0.6 &  0.5 & -0.4 \\
\hline
  60 & $7.28\cdot 10^{-3}$ & 0.13 & 0.809 &  1.8 &  1.9 &  0.8 &  0.0 &  0.9 & -0.7 & 0.0 & -0.2 & -0.1 &  0.4 &  1.2 & -0.1 \\
  60 & $4.30\cdot 10^{-3}$ & 0.22 & 0.970 &  2.0 &  2.2 &  0.9 &  0.0 &  1.4 & -0.2 & 0.0 & -0.2 &  0.5 &  0.7 &  1.1 & -0.3 \\
  60 & $3.05\cdot 10^{-3}$ & 0.31 & 1.123 &  2.3 &  1.8 &  1.1 & 0.0 & -0.6 & -0.1 & 0.0 & -0.1 & -0.4 &  0.8 &  0.9 & -0.3 \\
  60 & $2.37\cdot 10^{-3}$ & 0.40 & 1.183 &  2.7 &  1.6 &  1.3 & 0.0 & -0.2 & -0.1 &  0.0 &  0.7 & -0.3 & -0.2 &  0.3 & -0.4 \\
  60 & $1.97\cdot 10^{-3}$ & 0.48 & 1.135 &  3.3 &  2.2 &  1.5 & -0.1 &  0.6 & -0.5 &  0.3 &  0.7 & -0.9 & -0.2 & -0.5 & -0.4 \\
  60 & $1.69\cdot 10^{-3}$ & 0.56 & 1.277 &  3.6 &  2.9 &  1.8 & -0.3 & -1.0 & -0.3 & -0.1 &  1.0 &  0.7 &  1.4 &  0.7 & -0.4 \\
  60 & $1.50\cdot 10^{-3}$ & 0.63 & 1.417 &  4.4 &  3.0 &  2.3 & -0.3 &  1.3 & -0.4 &  0.2 &  0.6 &  0.4 &  0.8 & -0.3 & -0.5 \\
  60 & $1.37\cdot 10^{-3}$ & 0.69 & 1.300 &  5.2 &  4.4 &  3.1 & -1.0 &  1.8 & -0.7 &  0.8 &  1.5 & -1.2 &  0.5 & -0.4 & -0.4 \\
  60 & $1.26\cdot 10^{-3}$ & 0.75 & 1.446 &  5.6 &  4.8 &  3.1 & -0.9 & -1.4 & -2.1 & -0.4 & -1.8 & -0.7 &  0.4 & -1.1 & -0.5 \\
\hline
\multicolumn{16}{r}{{Continued on Next Page\ldots}} \\
\newpage
\multicolumn{16}{l}{{\ldots Continued from Previous Page}} \\
\hline
$Q^2$ & $x$ & $y$ & $\tilde{\sigma}^{e^+p}$ & $\delta_{stat}$ &$\delta_{sys}$ & $\delta_{unc}$ & $\delta_{\gamma p}$ & $\delta_{E_e}$ &$\delta_{E_{h}}$ &$\delta_{eID}$ & $\delta_{dx}$ & $\delta_{dy}$ & $\delta_{\rm MVD}$ &$\delta_{\rm CTD}$ &$\delta_{\rm diff}$\\
(GeV$^2$) & & & MER & (\%) & (\%) & (\%) & (\%) & (\%) & (\%) & (\%) & (\%) & (\%) & (\%) & (\%) & (\%) \\\hline
  80 & $9.71\cdot 10^{-3}$ & 0.13 & 0.751 &  2.1 &  1.7 &  1.0 &  0.0 &  0.8 & -0.7 &  0.0 &  0.7 &  0.2 & -0.3 &  0.6 & -0.1 \\
  80 & $5.74\cdot 10^{-3}$ & 0.22 & 0.947 &  2.3 &  1.9 &  1.1 &  0.0 &  1.3 & -0.1 &  0.1 & -0.7 & -0.2 & -0.5 &  0.3 & -0.2 \\
  80 & $4.07\cdot 10^{-3}$ & 0.31 & 1.067 &  2.7 &  1.4 &  1.3 &  0.0 &  0.4 & 0.0 &  0.0 &  0.4 & -0.1 & -0.1 &  0.3 & -0.3 \\
  80 & $3.16\cdot 10^{-3}$ & 0.40 & 1.106 &  3.2 &  2.2 &  1.5 & 0.0 & -1.4 & -0.1 &  0.3 & -0.3 &  0.5 &  0.1 &  0.0 & -0.3 \\
  80 & $2.63\cdot 10^{-3}$ & 0.48 & 1.170 &  3.7 &  2.5 &  1.8 & -0.2 & -0.7 & -0.3 & -0.5 & -1.0 &  0.3 & -0.6 & -0.8 & -0.4 \\
  80 & $2.25\cdot 10^{-3}$ & 0.56 & 1.217 &  4.2 &  3.4 &  2.2 & -0.4 & -0.7 & -0.6 &  0.7 & -1.1 & -0.4 & -1.4 & -1.3 & -0.3 \\
  80 & $2.00\cdot 10^{-3}$ & 0.63 & 1.246 &  5.5 &  3.9 &  2.9 & -0.5 &  0.7 & -1.3 &  0.6 & -0.2 & -0.4 &  1.3 &  1.2 & -0.4 \\
  80 & $1.83\cdot 10^{-3}$ & 0.69 & 1.274 &  6.1 &  4.5 &  3.7 & -1.0 &  0.3 & -0.4 &  0.8 & -1.2 &  0.7 & -0.5 &  1.5 & -0.5 \\
  80 & $1.68\cdot 10^{-3}$ & 0.75 & 1.461 &  6.3 &  5.3 &  3.6 & -0.8 & -2.4 & -1.8 & -1.2 &  0.4 & -0.1 & -0.6 & -1.6 & -0.6 \\
\hline
 110 & $1.33\cdot 10^{-2}$ & 0.13 & 0.730 &  2.4 &  2.1 &  1.1 &  0.0 &  0.8 & -0.6 &  0.1 & -0.7 &  0.2 &  1.0 &  0.7 & -0.1 \\
 110 & $7.89\cdot 10^{-3}$ & 0.22 & 0.835 &  2.8 &  2.3 &  1.3 &  0.0 &  1.4 & -0.2 &  0.1 &  1.1 &  0.6 & -0.2 & -0.4 & -0.1 \\
 110 & $5.60\cdot 10^{-3}$ & 0.31 & 0.971 &  3.2 &  1.7 &  1.5 & 0.0 &  0.4 & -0.1 & -0.1 &  0.2 &  0.4 &  0.1 & -0.3 & -0.2 \\
 110 & $4.34\cdot 10^{-3}$ & 0.40 & 1.078 &  3.8 &  2.5 &  1.8 & 0.0 & -0.6 & -0.1 &  0.6 & -0.6 & -0.7 & -0.5 & -1.1 & -0.3 \\
 110 & $3.62\cdot 10^{-3}$ & 0.48 & 1.116 &  4.4 &  3.5 &  2.3 & -0.2 & -2.3 & 0.0 & 0.0 & -0.3 & -0.3 & -1.1 &  0.8 & -0.4 \\
 110 & $3.10\cdot 10^{-3}$ & 0.56 & 1.192 &  5.0 &  4.5 &  2.7 & -0.5 & -1.0 & -0.4 & -1.5 &  1.3 & -0.4 &  0.6 &  2.7 & -0.5 \\
 110 & $2.75\cdot 10^{-3}$ & 0.63 & 1.127 &  6.6 &  4.0 &  3.4 & -0.4 &  1.0 & -0.5 &  0.4 & -0.9 &  0.7 &  1.2 &  0.3 & -0.5 \\
 110 & $2.51\cdot 10^{-3}$ & 0.69 & 1.174 &  7.7 &  6.4 &  4.1 & -0.7 & -2.7 & -1.1 &  0.4 & -0.6 & -0.4 & -2.4 &  3.0 & -0.2 \\
 110 & $2.31\cdot 10^{-3}$ & 0.75 & 1.354 & 10.9 &  9.1 &  6.2 & -1.0 &  4.4 & -1.6 &  0.4 &  3.3 &  0.3 & -1.6 & -2.7 & -0.6 \\
\hline
\multicolumn{16}{c}{\parbox{1.5\LTcapwidth}{}}\\
\multicolumn{16}{c}{\parbox{1.5\LTcapwidth}{
\normalsize{\bf Table 2: }\it
The reduced cross section, $\tilde{\sigma}$, for the reaction $e^+p\rightarrow e^+X$ at $\sqrt{s}=251$\gev.  
Further details as described in caption of Table~\ref{tab:sigHER}.
}}
\label{tab:sigMER}
\end{longtable}
\end{center}

\newpage

\begin{center}
\scriptsize
\begin{longtable}{|r|c|c||c|r|r||r|r|r|r|r|r|r|r|r|r|}
\hline
$Q^2$ & $x$ & $y$ & $\tilde{\sigma}^{e^+p}$ & $\delta_{stat}$ &$\delta_{sys}$ & $\delta_{unc}$ & $\delta_{\gamma p}$ & $\delta_{E_e}$ &$\delta_{E_{h}}$ &$\delta_{eID}$ & $\delta_{dx}$ & $\delta_{dy}$ & $\delta_{\rm MVD}$ &$\delta_{\rm CTD}$ &$\delta_{\rm diff}$\\
(GeV$^2$) & & & LER & (\%) & (\%) & (\%) & (\%) & (\%) & (\%) & (\%) & (\%) & (\%) & (\%) & (\%) & (\%) \\\hline
  24 & $3.64\cdot 10^{-3}$ & 0.13 & 0.864 &  1.8 &  2.7 &  0.9 &  0.0 &  1.2 & -0.8 &  0.0 &  0.7 &  0.7 & -1.2 &  1.4 & -0.1 \\
  24 & $2.15\cdot 10^{-3}$ & 0.22 & 1.043 &  1.6 &  2.8 &  0.8 &  0.0 &  1.7 & -0.2 & 0.0 & -0.6 &  0.3 & -0.6 &  1.8 & -0.2 \\
  24 & $1.53\cdot 10^{-3}$ & 0.31 & 1.136 &  1.5 &  2.0 &  0.7 & 0.0 & -0.4 & -0.2 &  0.0 & -0.4 & -0.3 & -1.3 &  1.2 & -0.4 \\
  24 & $1.18\cdot 10^{-3}$ & 0.40 & 1.184 &  1.6 &  2.0 &  0.7 & -0.1 & -0.7 & -0.3 &  0.2 & -1.0 &  0.3 & -0.6 &  1.2 & -0.4 \\
  24 & $9.86\cdot 10^{-4}$ & 0.48 & 1.195 &  1.7 &  2.7 &  0.8 & -0.2 & -0.2 & -0.4 &  0.1 & -0.3 & -0.2 &  1.4 &  2.1 & -0.4 \\
  24 & $8.45\cdot 10^{-4}$ & 0.56 & 1.260 &  1.7 &  2.0 &  0.9 & -0.3 &  0.7 & -0.7 &  0.4 & -0.8 & -0.2 & -0.8 &  0.7 & -0.5 \\
  24 & $7.51\cdot 10^{-4}$ & 0.63 & 1.256 &  2.2 &  3.1 &  1.3 & -0.8 &  0.3 & -1.5 & -0.7 & -0.3 &  0.5 & -1.8 & -0.8 & -0.5 \\
  24 & $6.86\cdot 10^{-4}$ & 0.69 & 1.260 &  2.5 &  3.5 &  1.6 & -1.3 &  1.3 & -1.9 &  0.3 & -0.3 &  0.5 & -0.4 &  1.3 & -0.6 \\
  24 & $6.31\cdot 10^{-4}$ & 0.75 & 1.247 &  2.9 &  5.6 &  2.1 & -2.0 &  0.6 & -4.1 &  0.8 & -1.1 & -0.9 &  1.2 &  0.6 & -0.6 \\
\hline
  32 & $4.85\cdot 10^{-3}$ & 0.13 & 0.848 &  1.3 &  1.9 &  0.6 &  0.0 &  1.1 & -0.8 & 0.0 & -0.4 & -0.2 & -0.6 &  1.0 & -0.1 \\
  32 & $2.87\cdot 10^{-3}$ & 0.22 & 0.977 &  1.3 &  2.1 &  0.6 & 0.0 &  1.6 & -0.2 & 0.0 &  0.2 &  0.3 & -0.5 &  1.1 & -0.2 \\
  32 & $2.04\cdot 10^{-3}$ & 0.31 & 1.083 &  1.4 &  1.4 &  0.6 & 0.0 & -0.4 & -0.2 &  0.1 & -0.5 &  0.3 & -0.1 &  0.9 & -0.3 \\
  32 & $1.58\cdot 10^{-3}$ & 0.40 & 1.153 &  1.6 &  1.2 &  0.7 & -0.1 & -0.5 & -0.2 & 0.0 & -0.2 &  0.2 & -0.1 &  0.7 & -0.3 \\
  32 & $1.31\cdot 10^{-3}$ & 0.48 & 1.216 &  1.8 &  1.7 &  0.8 & -0.1 & -0.6 & -0.3 &  0.1 & -0.3 & -0.3 & -0.7 &  0.8 & -0.4 \\
  32 & $1.13\cdot 10^{-3}$ & 0.56 & 1.249 &  2.0 &  1.8 &  1.0 & -0.3 &  0.8 & -0.7 & -0.5 &  0.2 &  0.3 & -0.5 &  0.4 & -0.4 \\
  32 & $1.00\cdot 10^{-3}$ & 0.63 & 1.252 &  2.6 &  2.3 &  1.5 & -0.7 & -0.3 & -1.1 & -0.5 &  0.4 & -0.3 &  0.9 & -0.1 & -0.4 \\
  32 & $9.15\cdot 10^{-4}$ & 0.69 & 1.387 &  2.8 &  3.5 &  1.7 & -0.9 & -0.1 & -2.1 & -0.7 &  0.5 & -0.2 &  1.1 &  1.3 & -0.5 \\
  32 & $8.41\cdot 10^{-4}$ & 0.75 & 1.291 &  3.3 &  4.0 &  2.3 & -1.7 &  0.3 & -2.6 &  0.6 & -0.3 & -0.3 & -0.3 & -0.4 & -0.6 \\
\hline
  45 & $6.83\cdot 10^{-3}$ & 0.13 & 0.803 &  1.2 &  1.7 &  0.5 & 0.0 &  0.8 & -0.7 &  0.0 &  0.2 &  0.1 &  0.5 &  1.1 & -0.1 \\
  45 & $4.03\cdot 10^{-3}$ & 0.22 & 0.942 &  1.3 &  2.1 &  0.6 &  0.0 &  1.5 & -0.1 &  0.1 & -0.5 & -0.2 &  0.4 &  1.2 & -0.2 \\
  45 & $2.86\cdot 10^{-3}$ & 0.31 & 1.057 &  1.5 &  1.1 &  0.7 & 0.0 & -0.1 & -0.1 &  0.1 & -0.4 & -0.1 &  0.1 &  0.7 & -0.3 \\
  45 & $2.22\cdot 10^{-3}$ & 0.40 & 1.107 &  1.7 &  1.5 &  0.8 & -0.1 & -0.4 & -0.2 & 0.0 & -0.3 & -0.3 & -0.7 &  0.7 & -0.3 \\
  45 & $1.85\cdot 10^{-3}$ & 0.48 & 1.105 &  2.0 &  1.4 &  1.0 & -0.2 & -0.4 & -0.4 &  0.2 & -0.3 & -0.1 &  0.4 &  0.5 & -0.4 \\
  45 & $1.58\cdot 10^{-3}$ & 0.56 & 1.260 &  2.2 &  2.3 &  1.2 & -0.3 &  0.1 & -0.5 &  1.0 & -0.6 &  0.2 & -1.0 & -0.8 & -0.4 \\
  45 & $1.41\cdot 10^{-3}$ & 0.63 & 1.293 &  2.8 &  2.4 &  1.4 & -0.3 & -0.7 & -1.0 & -0.9 &  0.7 &  0.3 & -0.8 &  0.4 & -0.4 \\
  45 & $1.29\cdot 10^{-3}$ & 0.69 & 1.227 &  3.2 &  3.0 &  1.9 & -0.8 &  0.6 & -1.5 & -0.3 &  0.7 &  0.1 & -0.4 &  0.9 & -0.5 \\
  45 & $1.18\cdot 10^{-3}$ & 0.75 & 1.228 &  3.7 &  4.5 &  2.4 & -1.4 & -2.1 & -2.0 & -0.5 & -0.3 &  0.6 & -1.0 & -1.2 & -0.6 \\
\hline
  60 & $9.10\cdot 10^{-3}$ & 0.13 & 0.746 &  1.3 &  1.9 &  0.6 &  0.0 &  1.0 & -0.7 &  0.1 & -0.6 &  0.3 &  0.5 &  1.0 & -0.1 \\
  60 & $5.38\cdot 10^{-3}$ & 0.22 & 0.868 &  1.5 &  2.6 &  0.7 & 0.0 &  1.4 & -0.2 & 0.0 & -0.3 &  0.3 &  1.2 &  1.7 & -0.2 \\
  60 & $3.82\cdot 10^{-3}$ & 0.31 & 0.994 &  1.8 &  1.1 &  0.8 &  0.0 &  0.2 & -0.1 &  0.1 &  0.1 &  0.4 &  0.3 &  0.5 & -0.3 \\
  60 & $2.96\cdot 10^{-3}$ & 0.40 & 1.068 &  2.1 &  1.5 &  0.9 & 0.0 & -0.8 & -0.2 & -0.1 & -0.1 & -0.1 &  0.3 &  0.6 & -0.3 \\
  60 & $2.46\cdot 10^{-3}$ & 0.48 & 1.112 &  2.4 &  1.6 &  1.2 & -0.2 &  0.3 & -0.2 &  0.3 &  0.6 &  0.2 & -0.2 &  0.5 & -0.3 \\
  60 & $2.11\cdot 10^{-3}$ & 0.56 & 1.126 &  2.7 &  2.3 &  1.3 & -0.3 & -0.2 & -0.5 &  0.6 &  1.1 &  0.1 & -0.3 &  1.1 & -0.4 \\
  60 & $1.88\cdot 10^{-3}$ & 0.63 & 1.215 &  3.4 &  3.4 &  1.8 & -0.4 &  1.5 & -0.8 &  0.3 &  1.8 & -0.2 &  0.6 &  1.0 & -0.4 \\
  60 & $1.71\cdot 10^{-3}$ & 0.69 & 1.290 &  3.7 &  3.2 &  2.2 & -0.8 &  0.9 & -1.6 & -0.4 & -0.3 &  0.5 &  0.7 & -0.7 & -0.5 \\
  60 & $1.58\cdot 10^{-3}$ & 0.75 & 1.221 &  4.2 &  3.7 &  2.4 & -0.9 &  1.0 & -1.9 &  0.7 & -1.0 &  0.3 &  0.4 & -1.0 & -0.4 \\
\hline
\multicolumn{16}{r}{{Continued on Next Page\ldots}} \\
\newpage
\multicolumn{16}{l}{{\ldots Continued from Previous Page}} \\
\hline
$Q^2$ & $x$ & $y$ & $\tilde{\sigma}^{e^+p}$ & $\delta_{stat}$ &$\delta_{sys}$ & $\delta_{unc}$ & $\delta_{\gamma p}$ & $\delta_{E_e}$ &$\delta_{E_{h}}$ &$\delta_{eID}$ & $\delta_{dx}$ & $\delta_{dy}$ & $\delta_{\rm MVD}$ &$\delta_{\rm CTD}$ &$\delta_{\rm diff}$\\
(GeV$^2$) & & & LER & (\%) & (\%) & (\%) & (\%) & (\%) & (\%) & (\%) & (\%) & (\%) & (\%) & (\%) & (\%) \\\hline
  80 & $1.21\cdot 10^{-2}$ & 0.13 & 0.708 &  1.6 &  1.6 &  0.7 & 0.0 &  0.9 & -0.6 &  0.0 & -0.4 & -0.2 & -0.6 &  0.4 & -0.1 \\
  80 & $7.17\cdot 10^{-3}$ & 0.22 & 0.882 &  1.7 &  1.8 &  0.8 &  0.0 &  1.5 & -0.2 &  0.1 & -0.3 & -0.2 & -0.4 & -0.2 & -0.1 \\
  80 & $5.09\cdot 10^{-3}$ & 0.31 & 0.954 &  2.0 &  1.1 &  0.9 & 0.0 & -0.2 & -0.1 &  0.1 & -0.2 & -0.4 & -0.3 &  0.3 & -0.2 \\
  80 & $3.94\cdot 10^{-3}$ & 0.40 & 1.020 &  2.4 &  2.2 &  1.1 & 0.0 & -1.1 & -0.1 &  0.1 & -0.2 & -0.5 & -1.0 & -1.0 & -0.3 \\
  80 & $3.29\cdot 10^{-3}$ & 0.48 & 1.076 &  2.8 &  1.9 &  1.4 & -0.2 & -0.6 & -0.1 &  0.1 &  0.5 & -0.5 &  0.3 &  0.6 & -0.4 \\
  80 & $2.82\cdot 10^{-3}$ & 0.56 & 1.101 &  3.1 &  3.8 &  1.6 & -0.3 & -1.1 & -0.4 &  1.0 &  0.2 &  0.4 &  2.3 &  2.0 & -0.3 \\
  80 & $2.50\cdot 10^{-3}$ & 0.63 & 1.113 &  4.0 &  3.2 &  2.0 & -0.3 & -0.6 & -0.7 &  0.6 & -1.0 & -0.3 & -0.5 & -1.7 & -0.4 \\
  80 & $2.29\cdot 10^{-3}$ & 0.69 & 1.074 &  4.8 &  5.2 &  3.4 & -1.4 &  1.6 & -1.7 &  1.0 & -1.1 & -0.5 &  1.0 &  2.1 & -0.4 \\
  80 & $2.10\cdot 10^{-3}$ & 0.75 & 1.241 &  4.8 &  4.9 &  2.9 & -0.9 &  1.9 & -1.1 & -1.6 & -1.5 & -1.3 & -0.3 & -1.8 & -0.5 \\
\hline
 110 & $1.67\cdot 10^{-2}$ & 0.13 & 0.666 &  1.8 &  1.4 &  0.8 &  0.0 &  0.7 & -0.7 &  0.1 & -0.4 &  0.1 & -0.4 & -0.3 & -0.1 \\
 110 & $9.86\cdot 10^{-3}$ & 0.22 & 0.812 &  2.0 &  1.9 &  0.9 &  0.0 &  1.4 & -0.1 & -0.2 &  0.1 & -0.5 & -0.4 &  0.2 & -0.2 \\
 110 & $7.00\cdot 10^{-3}$ & 0.31 & 0.853 &  2.4 &  1.8 &  1.1 & 0.0 &  0.2 & -0.1 & -0.1 &  0.6 &  0.3 &  0.9 &  0.9 & -0.2 \\
 110 & $5.42\cdot 10^{-3}$ & 0.40 & 0.964 &  2.9 &  2.4 &  1.2 & 0.0 & -1.4 & -0.1 &  0.1 &  0.8 &  0.6 &  1.0 &  0.4 & -0.3 \\
 110 & $4.52\cdot 10^{-3}$ & 0.48 & 1.053 &  3.2 &  1.9 &  1.5 & -0.1 & -0.5 & -0.2 &  0.5 & -0.6 &  0.1 &  0.5 & -0.4 & -0.3 \\
 110 & $3.87\cdot 10^{-3}$ & 0.56 & 1.026 &  3.8 &  3.1 &  1.7 & -0.2 &  0.4 & -0.3 & -1.8 & -0.9 & -0.2 &  0.6 & -1.2 & -0.4 \\
 110 & $3.44\cdot 10^{-3}$ & 0.63 & 1.086 &  4.9 &  3.9 &  2.8 & -0.7 & -0.9 & -0.6 & -0.4 & -1.2 &  0.4 &  1.7 & -0.9 & -0.5 \\
 110 & $3.14\cdot 10^{-3}$ & 0.69 & 1.141 &  5.4 &  4.3 &  3.2 & -0.7 &  0.5 & -0.8 & -0.1 &  1.3 &  0.5 &  1.9 &  1.2 & -0.4 \\
 110 & $2.89\cdot 10^{-3}$ & 0.75 & 0.916 &  9.4 &  7.6 &  5.2 & -1.0 & -1.8 & -1.4 &  1.5 &  1.7 &  3.0 & -1.1 & -2.8 & -0.4 \\
\hline
\multicolumn{16}{c}{\parbox{1.5\LTcapwidth}{}}\\
\multicolumn{16}{c}{\parbox{1.5\LTcapwidth}{
\normalsize{\bf Table 3: }\it
The reduced cross section, $\tilde{\sigma}$, for the reaction $e^+p\rightarrow e^+X$ at $\sqrt{s}=225$\gev. 
Further details as described in caption of Table~\ref{tab:sigHER}.
}}
\label{tab:sigLER}
\end{longtable}
\end{center}

\newpage 

\begin{table}
\begin{center}
\footnotesize
\begin{tabular}{|r|c||r|r||r|r|}
\hline
$Q^2$     & $x$ & $F_L^{(1)}$ & $F_L^{(2)}$ & $F_2^{(1)}$ & $F_2^{(2)}$ \\
(GeV$^2$) &     &             &             &             &             \\ 
\hline
24 & $6.67\cdot 10^{-4}$ & $0.29_{-0.11}^{+0.11}$ & $0.31_{-0.08}^{+0.14}$ & $1.403_{-0.025}^{+0.025}$ & $1.404_{-0.022}^{+0.029}$ \\
24 & $8.16\cdot 10^{-4}$ & $0.25_{-0.13}^{+0.13}$ & $0.27_{-0.11}^{+0.14}$ & $1.340_{-0.029}^{+0.029}$ & $1.341_{-0.028}^{+0.030}$ \\
24 & $1.08\cdot 10^{-3}$ & $0.46_{-0.22}^{+0.22}$ & $0.47_{-0.14}^{+0.29}$ & $1.260_{-0.028}^{+0.028}$ & $1.261_{-0.020}^{+0.035}$ \\
\hline
32 & $8.89\cdot 10^{-4}$ & $0.18_{-0.11}^{+0.11}$ & $0.20_{-0.07}^{+0.14}$ & $1.397_{-0.023}^{+0.023}$ & $1.398_{-0.018}^{+0.028}$ \\
32 & $1.09\cdot 10^{-3}$ & $0.26_{-0.13}^{+0.13}$ & $0.27_{-0.10}^{+0.14}$ & $1.324_{-0.021}^{+0.021}$ & $1.324_{-0.019}^{+0.023}$ \\
32 & $1.43\cdot 10^{-3}$ & $0.27_{-0.19}^{+0.19}$ & $0.28_{-0.13}^{+0.22}$ & $1.229_{-0.023}^{+0.023}$ & $1.229_{-0.016}^{+0.027}$ \\
\hline
45 & $1.25\cdot 10^{-3}$ & $0.14_{-0.11}^{+0.11}$ & $0.15_{-0.07}^{+0.12}$ & $1.324_{-0.024}^{+0.024}$ & $1.324_{-0.018}^{+0.029}$ \\
45 & $1.53\cdot 10^{-3}$ & $-0.11_{-0.13}^{+0.13}$ & $0.00_{-0.00}^{+0.10}$ & $1.233_{-0.022}^{+0.022}$ & $1.246_{-0.010}^{+0.027}$ \\
45 & $2.02\cdot 10^{-3}$ & $0.37_{-0.19}^{+0.19}$ & $0.38_{-0.13}^{+0.24}$ & $1.173_{-0.022}^{+0.022}$ & $1.173_{-0.016}^{+0.028}$ \\
\hline
60 & $1.67\cdot 10^{-3}$ & $0.16_{-0.12}^{+0.12}$ & $0.17_{-0.09}^{+0.13}$ & $1.326_{-0.029}^{+0.029}$ & $1.326_{-0.021}^{+0.034}$ \\
60 & $2.04\cdot 10^{-3}$ & $0.19_{-0.15}^{+0.15}$ & $0.21_{-0.11}^{+0.17}$ & $1.211_{-0.025}^{+0.025}$ & $1.211_{-0.017}^{+0.030}$ \\
60 & $2.69\cdot 10^{-3}$ & $0.27_{-0.21}^{+0.21}$ & $0.28_{-0.14}^{+0.24}$ & $1.145_{-0.024}^{+0.024}$ & $1.145_{-0.015}^{+0.030}$ \\
\hline
80 & $2.22\cdot 10^{-3}$ & $0.18_{-0.13}^{+0.13}$ & $0.19_{-0.09}^{+0.15}$ & $1.256_{-0.029}^{+0.029}$ & $1.256_{-0.022}^{+0.034}$ \\
80 & $2.72\cdot 10^{-3}$ & $0.38_{-0.17}^{+0.17}$ & $0.39_{-0.14}^{+0.19}$ & $1.213_{-0.027}^{+0.027}$ & $1.214_{-0.022}^{+0.032}$ \\
80 & $3.59\cdot 10^{-3}$ & $0.12_{-0.22}^{+0.22}$ & $0.13_{-0.13}^{+0.19}$ & $1.082_{-0.022}^{+0.022}$ & $1.081_{-0.010}^{+0.029}$ \\
\hline
110 & $3.06\cdot 10^{-3}$ & $0.33_{-0.15}^{+0.15}$ & $0.34_{-0.12}^{+0.18}$ & $1.254_{-0.033}^{+0.033}$ & $1.255_{-0.028}^{+0.039}$ \\
110 & $3.74\cdot 10^{-3}$ & $0.31_{-0.18}^{+0.18}$ & $0.33_{-0.15}^{+0.21}$ & $1.136_{-0.029}^{+0.029}$ & $1.138_{-0.023}^{+0.035}$ \\
110 & $4.93\cdot 10^{-3}$ & $0.01_{-0.25}^{+0.25}$ & $0.03_{-0.03}^{+0.25}$ & $1.022_{-0.026}^{+0.026}$ & $1.022_{-0.006}^{+0.037}$ \\
\hline
\end{tabular}
\caption{The measured values of $F_L$ and $F_2$ at 18 separate $(x,Q^2)$ points.  
The quoted uncertainties 
include both the statistical and systematic sources, 
whereas a $\pm2.5$\% correlated normalisation uncertainty is not included on the $F_L$ 
and $F_2$ values. The $(1)$ superscript indicates an unconstrained fit whereas the $(2)$ 
superscript indicates that the constraints $F_2\ge 0$ and $0\le F_L\le F_2$ 
were enforced by the prior.}
\label{tab:FL_F2}
\end{center}
\end{table}

\begin{table}
\begin{center}
\footnotesize
\begin{tabular}{|r||r|r|}
\hline
$Q^2$     & $F_L^{(1)}$ & $F_L^{(2)}$ \\
(GeV$^2$) &             &             \\
\hline
24  & $0.30_{-0.09}^{+0.09}$ & $0.30_{-0.07}^{+0.10}$ \\
\hline
32  & $0.22_{-0.09}^{+0.09}$ & $0.22_{-0.07}^{+0.09}$ \\
\hline
45  & $0.10_{-0.08}^{+0.08}$ & $0.10_{-0.07}^{+0.07}$ \\
\hline
60  & $0.18_{-0.09}^{+0.09}$ & $0.18_{-0.08}^{+0.10}$ \\
\hline
80  & $0.24_{-0.10}^{+0.10}$ & $0.23_{-0.09}^{+0.11}$ \\
\hline
110 & $0.28_{-0.12}^{+0.12}$ & $0.28_{-0.11}^{+0.13}$ \\
\hline
\end{tabular}
\caption{The single values of $F_L$ extracted in each $Q^2$ bin. The quantities are quoted such that $Q^2/x$ was constant for each value, 
which corresponds to \mbox{$y=0.71$} for $\sqrt{s}=225$~GeV. 
The quoted uncertainties 
include both the statistical and systematic sources, although a $\pm2.5$\% normalisation uncertainty is not included.  The $(1)$ superscript indicates an unconstrained fit whereas the $(2)$ superscript indicates that the constraints $F_2\ge 0$ and $0\le F_L\le F_2$ were 
enforced in the prior.}
\label{tab:FL_Q2}
\end{center}
\end{table}

\begin{table}
\begin{center}
\footnotesize
\begin{tabular}{|r||r|r|}
\hline
$Q^2$     & $R^{(1)}$ & $R^{(2)}$  \\
(GeV$^2$) &           &             \\
\hline
24  & $0.27_{-0.07}^{+0.14}$ & $0.27_{-0.06}^{+0.15}$  \\
\hline
32  & $0.18_{-0.05}^{+0.12}$ & $0.18_{-0.05}^{+0.12}$  \\
\hline
45  & $0.08_{-0.05}^{+0.11}$ & $0.08_{-0.04}^{+0.10}$  \\
\hline
60  & $0.16_{-0.07}^{+0.13}$ & $0.16_{-0.07}^{+0.13}$  \\
\hline
80  & $0.23_{-0.09}^{+0.17}$ & $0.23_{-0.08}^{+0.17}$  \\
\hline
110 & $0.29_{-0.12}^{+0.23}$ & $0.29_{-0.12}^{+0.22}$  \\ 
\hline
\end{tabular}
\caption{The single values of $R$ extracted in each $Q^2$ bin. Other
  details as in the caption to Table~\ref{tab:FL_Q2}, although no
  additional $\pm2.5$\% normalisation uncertainty need be included.}
\label{tab:R_Q2}
\end{center}
\end{table}


%
%
\end{document}